\documentclass[useAMS,fleqn,usenatbib]{mnras}
\usepackage[T1]{fontenc}
\usepackage{ae,aecompl}
\usepackage{bethmacros}
\usepackage{graphicx}
\usepackage{amssymb}
\usepackage{amsmath}
\usepackage{placeins}
\usepackage{times}

\def\spose#1{\hbox to 0pt{#1\hss}}
\def\lta{\mathrel{\spose{\lower 3pt\hbox{$\mathchar"218$}}
     \raise 2.0pt\hbox{$\mathchar"13C$}}}
\def\gta{\mathrel{\spose{\lower 3pt\hbox{$\mathchar"218$}}
     \raise 2.0pt\hbox{$\mathchar"13E$}}}

\title[Supernova input rates]{Uncertainties in supernova input rates
drive qualitative differences in simulations of galaxy evolution}

\author[B.W. Keller and J.M.D. Kruijssen]{Benjamin W. Keller\thanks{Email: benjamin.keller `at'
uni-heidelberg.de}  and J. M. Diederik Kruijssen
\vspace*{6pt}\\
Astronomisches Rechen-Institut, Zentrum f{\"u}r Astronomie der Universit\"at
Heidelberg, M{\"o}nchofstra{\ss}e 12-14, D-69120 Heidelberg, Germany}

\begin{document}
\maketitle
\label{firstpage}
\begin{abstract} 
    Feedback from core collapse supernovae (SNe), the final stage of evolution
    of massive stars, is a key element in galaxy formation theory.  The energy
    budget of SN feedback, as well as the duration over which SNe occur, are
    constrained by stellar lifetime models and the minimum mass star that ends
    its life as a SN.  Simplifying approximations for this SN rate are
    ubiquitous in simulation studies. We show here how the choice of SN budget
    and timings ($t_0$ for the delay between star formation and the first SN,
    $\tau_{\rm SN}$ for the duration of SN injection, and the minimum SN progenitor mass)
    drive changes in the regulation of star formation and outflow launching.
    Extremely long delays for instantaneous injection of SN energy $(t_0\gg 20\Myr)$
    reduces star formation and drive stronger outflows compared smaller delays.
    This effect is primarily driven by enhanced clustering of young stars.
    With continuous injection of energy, longer SN durations results in a larger
    fraction of SN energy deposited in low ambient gas densities, where cooling
    losses are lower.  This is effect is particularly when driven by the choice
    of the minimum SN progenitor mass, which also sets the total SN energy
    budget.  These underlying uncertainties mean that despite advances in the
    sub-grid modeling of SN feedback, serious difficulties in constraining the
    strength of SN feedback remain.  We recommend future simulations use
    realistic SN injection durations, and bound their results using SN energy
    budgets and durations for minimum SN progenitors of $7\Msun$ and $9\Msun$.
\end{abstract}

\begin{keywords}
    galaxies : formation --  galaxies : evolution -- supernovae: general -- 
    galaxies : star formation -- methods : numerical
\end{keywords}

\section{Introduction}
\label{introduction}
Supernovae (SNe) are important components of any theory of galaxy formation, in
both (semi-)analytic models and hydrodynamic simulations.  The energy and
momentum injected by SNe stirs turbulence in the interstellar medium (ISM)
\citep{Joung2006}, drives the formation of a multiphase ISM \citep{McKee1977},
and can launch galaxy-scale winds and outflows \citep{Larson1974}.  Modeling
SNe in galaxy formation simulations has proven to be non-trivial. The failure of
SNe to regulate star formation in early simulations
\citep{Katz1992,Scannapieco2012} has led to significant effort to develop
sub-grid treatments of SNe in simulations where the early, small-scale evolution
of single SN remnants are unresolved
\citep{Thacker2000,Springel2003,Stinson2006,DallaVecchia2012,Keller2014}.

Any model for SN feedback must include choices for:
\begin{itemize}
    \item The amount of energy/momentum injected ({\it the SN budget}).
    \item The timescale over which the injection occurs ({\it the SN rate}).
    \item A method to counteract numerical errors from the resolution of
        spatial/temporal discretization ({\it the sub-grid model}).
\end{itemize}
The most technically challenging of these is the final, and numerous studies
have proposed different ways to properly treat the energy injected by feedback
without dealing with ``overcooling'' and to achieve resolution convergence.
Despite the great sophistication and success of many of these models, less
consideration has gone into how the SN input rates and timescales change the
impact of SN feedback on galaxy evolution. Instead, many have used very simplified
assumptions about the SN budget and rate.

The ``overcooling problem'' was identified early on, in some of the first
cosmological hydrodynamic simulations that included SN feedback
\citep[e.g.][]{Katz1992}.  In these simulations, SN energy was deposited into
resolution elements that contained many orders of magnitude more mass than the
SN ejecta.  As such, two related problems arose.  The first was that the Sedov
radius for individual SNe was also many orders of magnitude smaller than the
hydrodynamical resolution of the simulations.  By failing to resolve the
energy-conserving Sedov-Taylor phase \citep{Taylor1950}, the amount of
shock-heated gas becomes a function of numerical resolution alone.  On top of
this, the generation of momentum by this phase and the later pressure-driven
snowplow (when the swept-up shell can cool, but the hot bubble remains
adiabatic) is completely lost if the resolution isn't smaller than the Sedov
radius \citep{Sedov1959}. This minimum required resolution can be as small as
$\sim 1\pc$ in a typical galaxy \citep{Hu2019}.  The second problem, related to
this, arises from the radiative cooling rate of the hot bubble.  If too much
mass is heated by SN feedback, the temperature of this bubble may be close to
the peak of the radiative cooling curve at $10^5\K$.  If this occurs, the
cooling time for the bubble will be artificially lowered, potentially by an
order of magnitude or more \citep{DallaVecchia2012}.  These two numerical
effects can lead to SNe radiating nearly all of their energy away before they
have a chance to either generate significant momentum or hot, buoyant gas that
rises into the galactic halo \citep{Keller2020a}.  As the resolution required to
avoid these problems is still beyond the reach of cosmological simulations, this
has motivated the focus on numerical sub-grid models to overcome overcooling.

A great deal of work has been done to study the effects of changing the
numerical method for coupling feedback energy and momentum to the ISM.  Many new
sub-grid feedback models that have been proposed in recent decades
\citep[e.g.][]{Navarro1993,Gerritsen1997,Springel2003,Scannapieco2006,Kimm2014,Keller2014,Smith2018}.
As a result of this growing number of models, studies have been undertaken to
try and quantify the effects of individual model parameters, as well as
the differences between models.  Two early attempts at this were
\citet{Thacker2000} and \citet{Kay2002}.  \citet{Thacker2000} showed that in
smoothed-particle hydrodynamics (SPH) simulations, the choice of how to smooth
feedback energy over a region can have significant effects on whether feedback
energy is lost to numerical overcooling. \citet{Kay2002} studied how kinetic
feedback and thermal feedback implemented using the \citet{Gerritsen1997}
delayed cooling model varies in effectiveness for different cooling delay times
and SN efficiencies.  Unsurprisingly, they found lower star formation rates
(SFRs) and cold gas fractions with higher feedback efficiencies.  For thermal
feedback with a cooling delay, they also found that star formation cold be
regulated only when unrealistically large delay times (approaching a Hubble
time) were used.  Subsequent studies have refined these results, as well as
showing how many of the issues of numerical overcooling are alleviated through
careful consideration of the unresolved evolution of feedback bubbles.
\citet{DallaVecchia2012} showed that the amount of mass heated by feedback was a
critical component to determining its effectiveness, while \citet{Kimm2014}
showed that attempting to match the terminal momentum of high-resolution studies
of SN bubbles could produce a resolution-insensitive model for momentum
injection via SN feedback.

The failure of the first generation of SN models also drove a push by simulators
to examine additional forms of ``early'' (pre-SNe) feedback.  Models have now
been developed that include stellar winds, radiation pressure, and ionization by
UV photons, in addition to SNe
\citep[e.g.][]{Stinson2013,Agertz2013,Hopkins2014}.  Beyond the additional
momentum and energy these mechanisms provide, they are believed to ``prime'' the
regions of SN detonation, reducing the cooling losses experienced by SNe
\citep{Rogers2013}.  Beyond these more traditional forms of stellar feedback,
significant work is now also being done to better understand the role of
high-energy cosmic rays in heating and stirring the ISM
\citep[e.g.][]{Jubelgas2008,Booth2013,Girichidis2016a}.  Despite the complex
array of potential sources of stellar feedback, there remains significant
uncertainty in how SNe alone impact the evolution of galaxies.

A comprehensive comparison of simulations including different feedback, star
formation, and hydrodynamic methods was undertaken by \citet{Scannapieco2012},
who found that the choice of feedback model (as well as the underlying physical
process driving that feedback) produces far greater variation in cosmological
galaxy evolution than differences in hydrodynamics method.  A more tightly
controlled study by \citet{Rosdahl2017} implemented five different sub-grid
models for feedback in the simulation code {\sc Ramses} \citep{Teyssier2002}.
These models were a simple thermal dump without any mechanism to control
overcooling, a stochastic thermal model based on \citet{DallaVecchia2012}; a
delayed cooling model similar to \citet{Gerritsen1997} first introduced in
\citet{Teyssier2013}; a kinetic feedback model introduced in \citet{Dubois2008};
and a mechanical feedback mechanism based on \citet{Kimm2014}.  Using the same
initial conditions, star formation method, and hydrodynamics solver allowed them
to control their study to look at the impact of the feedback sub-grid model
alone.  They found significant differences in the qualitative appearance of
their simulated galaxies, as well as quantitative changes in the star formation
and outflow rates, outflow properties, and temperature-density phase diagrams.

The total energy budget and the duration of SN feedback for a stellar population
depends both on stellar evolution and the initial mass function (IMF).  Stellar
evolution models can predict whether a star with a given mass and metallicity
will end its life as a core-collapse SN \citep[e.g.][]{Smartt2009}, the energy
released by that SN, and the lifetime of the star prior to SN detonation
\citep[e.g.][]{Leitherer1999,Ekstrom2012,Leitherer2014}.  By calculating the
range of stellar masses which produce SNe, we can use the IMF to determine the
SN energy budget and rate of a stellar population.  While there are constraints
on many of these pieces, there are still sufficiently many uncertainties that
the total amount of energy released by SNe cannot be well constrained with better
than $\sim50$ per cent uncertainty.  Because the lowest mass SN progenitors have
the longest lifetimes, this also leads to a corresponding uncertainty in the
duration over which SN feedback occurs, $\tau_{\rm SN}$.  This picture is
complicated by the fact that many simulation approaches simply ignore the
different lifetimes of stars, injecting an entire stellar population's SN budget
instantaneously \citep[e.g.][]{Crain2015,Rosdahl2017}.  The total energy budget
from one set of simulations to the next can vary by more than a factor of 2, and
injection timescales can vary by tens of $\Myr$ (see
Table~\ref{other_SN_rates}).  

The structure of this paper is as follows. First, we show how the SN injection rates
and energy budget can be calculated, and compare this to a brief survey of
injection rates used in the literature (Section~\ref{injection_rates}).  We then
introduce a set of simulations of an isolated Milky Way galaxy analogue that we
will use to study the effects of varying the SN injection rate and timescale
(Section~\ref{methods}).  With these simulations, we examine the impact of
instantaneously injecting all SN energy after varying delay times
(Section~\ref{delay_time}), comparing this to the effect of using a continuous
injection based on the stellar lifetimes of SN progenitors
(Section~\ref{continuous}).  We then show the importance of the minimum SN
progenitor mass (Section~\ref{SN_min_mass}).  Finally, we conclude with a
discussion of the implications of these results, and what is needed going
forward to build simulations that offer strong predictive power
(Section~\ref{discussion} and~\ref{conclusion}).
\begin{table*}
    \begin{tabular}{ccccc}
        \hline
        Simulations & $\mathcal{E}_{\rm SN} (10^{49} \erg/\Msun)$ & 
        $t_0 (\Myr)$ & $\tau_{\rm SN} (\Myr)$ & SN Rate Form\\
        \hline
        \hline
        \citet{Springel2003} & 0.400 & 0 & 0 & Instantaneous \\ 
        \citet{Dobbs2011} & 0.625 & 0 & 0 & Instantaneous \\
        \citet{Dubois2012} & 1.000 & 10 & 0 & Instantaneous \\
        \citet{Agertz2013} & 1.014 & 4.92 & 35.21&
        Equation~\ref{sn_rates_general} \\
        \citet{Ceverino2014} & 1.488 & 0 & 40 & Constant Rate\\
        EAGLE \citep{Crain2015} & 1.730 & 30 & 0 & Instantaneous \\
        AGORA Comparison  \citep{Kim2016} & $1.080$ & 5 & 0 & Instantaneous \\
        Snap, Crackle, Pop \citep{Rosdahl2017} & 2.000 & 5 & 0 & Instantaneous \\
        FIRE2 \citep{Hopkins2018b} & 1.060 & 3.4 & 34.13 &
        Equation~\ref{fire_rate} \\
        MUGS2 \citep{Keller2016} & 1.010 & 4.92 & 35.21 &
        Equation~\ref{sn_rates_general} \\ 
        \citet{Semenov2018} & 1.000 & 0 & 40 &Constant Rate \\
        FOGGIE \citep{Peeples2019} & 1.787 & $12t_{\rm dyn}$ & 0 & Instantaneous \\
        \hline
    \end{tabular}
    \caption{SN energy injection parameters for a set of simulations from the
    literature.  We show here the total specific energy injected through SNe
    $(\mathcal{E}_{\rm SN})$, the delay between star formation and the first SN
    event $(t_0)$, and timescale over which SNe occur $(\tau_{\rm SN})$.  As is
    clear, there is a spread in the total energy injected more than a factor of
    two, as well as a great variety in the SN delays and timescales.  Many
    simulations deposit SN energy as a single event, sometimes at the time of
    the first SN, sometimes at the midpoint of the distribution, or sometimes
    at time of the final SN.}
    \label{other_SN_rates}
\end{table*}

\section{A Survey of supernovae injection rates}
\label{injection_rates}
\begin{figure}
    \includegraphics[width=0.5\textwidth]{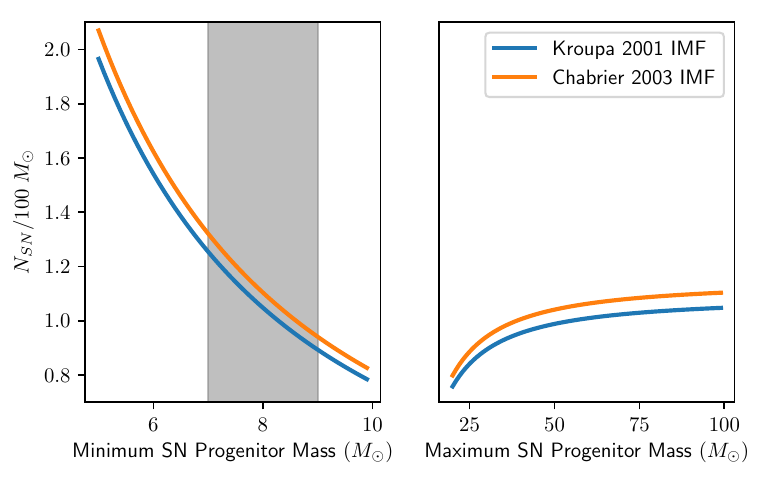}
    \caption{Variation in the number of SNe $(N_{\rm SN})$ occurring for a
    stellar population for different minimum and maximum progenitor masses.  The
    left panel show the change in $N_{\rm SN}$ for varying minimum mass with a
    maximum mass of $100\Msun$, while the right panel show the change with a
    minimum mass of $8\Msun$ and a varying maximum mass.  Blue curves use a
    \citet{Kroupa2001} IMF, while the orange curves use a \citet{Chabrier2003}
    IMF.  The vertical grey bar shows the best estimate minimum progenitor mass
    of $8\pm1\Msun$ \citep{Smartt2009}. The maximum progenitor mass is much more
    uncertain, and may span the entire range shown here. As is clear, the
    minimum progenitor mass has a more significant impact on the SN budget than
    the choice of maximum progenitor.}
    \label{imf_limits}
\end{figure}
\begin{figure}
    \includegraphics[width=0.5\textwidth]{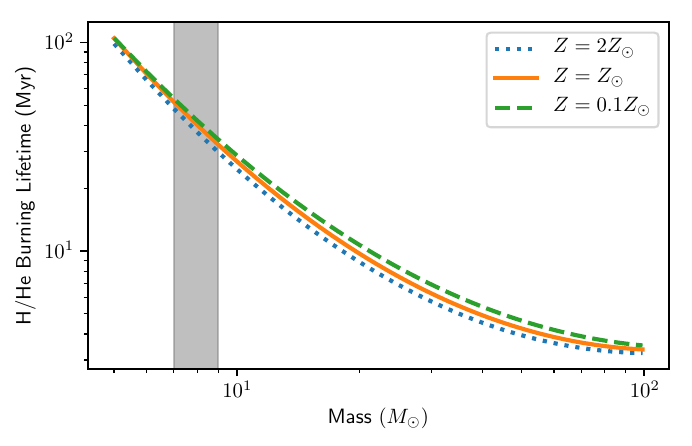}
    \caption{Hydrogen and Helium burning lifetimes as a function of stellar mass
    for stars with twice solar metallicity (blue dotted curve), solar
    metallicity (orange solid curve), and tenth solar metallicity (green dashed
    curve).  The vertical gray bar shows the best-estimate minimum progenitor
    mass for SNe ($8\pm1\Msun$).  As is clear, the mass of the star is a much
    stronger determinant for its lifetime than the star's metallicity.}
    \label{sn_times}
\end{figure}
The starting point for determining the energy budget of SNe produced by a simple
stellar population is the IMF \citep{Salpeter1955,Bastian2010,Offner2014}.  The typical 
form of the IMF is given as the number of stars in a logarithmic mass interval
$\Phi(M) = dN/d\log M$.  For high-mass stars, this is typically a power law
$\Phi(M) = A M^{-\Gamma}$, with the ``Salpeter slope'' being $\Gamma \sim 1.35$.
Naturally, the linear form of the IMF is related to the logarithmic form as
$\mathcal{X}(M) = dN/dM = \Phi(M)/(M\ln10)$.  Taking the first moment of this IMF form
allows us to determine the normalization for the IMF, such that the total mass
integrates to unity.  Unfortunately, if the IMF is a power-law, with a low-end
slope $\Gamma > 1$, it will diverge for the lower limits of integration as
$\lim_{M\to0}$.  This, along with better observations of low-mass stars, has led
to the development of newer IMF forms, consisting of multiple power law IMFs,
such as the commonly used \citet{Kroupa2001} IMF:
\begin{equation}
    \Phi(M) = A_k
    \begin{cases}
        25M^{0.7} & 0.01 \leq M/\Msun < 0.08 \\
        2M^{-0.3} & 0.08 \leq M/\Msun < 0.50 \\
        M^{-1.3} & 0.50 \leq M/\Msun < 100\\
    \end{cases}
    \label{kroupa2001_imf}
\end{equation}
An alternative form is to use a log-normal distribution with a power law tail,
as in the (also common) \citet{Chabrier2003} IMF.
\begin{equation}
    \Phi(M) = A_c
    \begin{cases}
        0.158\exp{(-\frac{(\log(M)+1.102)^2}{0.952})} & M/\Msun \leq 1.0\\
        0.28M^{-1.3} & M/\Msun > 1.0\\
    \end{cases}
    \label{chabrier2003_imf}
\end{equation}
With these forms, the lower mass limits no longer produce a divergent number of
low-mass stars.  None the less, we still wish to impose a lower limit on the
mass of stars for physical reasons, as we expect there to be a minimum mass for
core fusion in stars.  We also expect there to be an upper mass limit, somewhere
near the Eddington limit during the formation of massive stars.  These mass
limits are of course a matter of some debate, and can produce a small change in
the normalization of the IMF $(A$), on the order of 5 per cent for the most
extreme variations in both the lower and upper integration limits.  For this
paper, we will use a lower limit of $0.01\Msun$ and an upper limit of $100\Msun$
for calculations involving the IMF.  The normalization can be determined such
that that $\Phi(M)$ is given per unit mass by solving:
\begin{equation}
    \int_{0.01}^{100}M\mathcal{X}(M)dM =
    \int_{0.01}^{100}\frac{\Phi(M)}{\ln10}dM  = 1
\end{equation}
Together, this gives us a normalization for a \citet{Kroupa2001} IMF of $A_k =
0.487$ and for a Chabrier IMF $A_c = 0.512$.  The number of SNe that then
occur for a population is simply $\int\mathcal{X}MdM$ integrated over the initial
mass range of SN progenitors. 

Determining the number of SNe that occur per unit mass of stars formed is then
simply a question of choosing what masses of stars end their lives as a SN.
The choice of these limits is critical, as is shown in Figure~\ref{imf_limits}.
While the upper limit has less impact on the SN budget (because of the steep
Salpeter slope at high mass), changing the minimum progenitor mass from
$10\Msun$ to $6\Msun$ can change the total number of SNe occurring per unit mass
by a factor of $\sim2$, doubling the energy budget for SNe.  Converting the IMF
into an SN rate is simply a question of combining it with a model for stellar
lifetimes (as a function of their mass).  \citet{Raiteri1996} has produced a fit
for the Hydrogen and Helium burning lifetimes of stars computed by the Padova
group \citep{Alongi1993,Bressan1993,Bertelli1994}.  This simple log-quadratic
fit for stars with mass between $0.6M_\odot$ and $120M_\odot$ (well encompassing
the range of masses for SN progenitors) is given by
equation~\ref{raiteri_lifetimes}, 

Equation~\ref{raiteri_lifetimes} gives stellar lifetimes in years, as a function
of their metallicity $Z$ and their mass $M$.  
\begin{equation}
    \begin{split}
        \log (t_*) = a_0(Z) + a_1(Z)\log\left(\frac{M}{\Msun}\right) +
        a_2(Z)\log^2\left(\frac{M}{\Msun}\right), \\ 
        a_0 = 10.13 + 0.07547\log(Z) - 0.008084\log^2(Z), \\
        a_1 = -4.424 - 0.7939\log(Z) - 0.1187\log^2(Z), \\
        a_2 = 1.262 + 0.3385\log(Z) + 0.05417\log^2(Z) \\
    \end{split}
    \label{raiteri_lifetimes}
\end{equation}
The stellar
lifetimes are a strong function of their initial mass (as we might expect), but
this fit does have some dependence on stellar metallicity (as is shown in
Figure~\ref{sn_times}).  We can simply invert this function and combine it with
our IMF of choice to determine a SN rate as a function of time.  SN progenitors
are all massive enough that they fall in the Salpeter end of the IMF, with
$\Phi(M)\propto M^{-1.3}$, for equations~\ref{kroupa2001_imf}
and~\ref{chabrier2003_imf} resulting in a SN rate $dN_{\rm SN}/dt$ given by
\begin{equation}
    \log \frac{dN_{\rm SN}}{dt} = \log{A}+1.3\frac{a_1+\sqrt{a_1^2-4a_2(a_0-\log{t})}}{2a_2}
    \label{sn_rates_general}
\end{equation}
Despite this form, the actual rate for reasonable masses ($\sim5-100\Msun$) of
SN progenitors is relatively constant, with the first SN detonating at
$\sim3-5\Myr$, and the final SN detonating at $\sim30-70\Myr$.

A number of simulation codes and projects have used these rates directly to
stochastically seed SNe (beginning with the original work of
\citealt{Raiteri1996}), while others use the lifetimes to inform simpler models
for the SN rates.  An example of a fit to this relation (derived using
{\sc STARBURST99} \citep{Leitherer1999} simulations) is the function used in the FIRE
simulations for the SN rate:
\begin{equation}
    N_{\rm SN}/\Msun = 
    \begin{cases}
        0 & t_6 < 3.4\\
        (5.408t_6-18.39)\times10^{-4} & 3.4 < t_6 < 10.37 \\
        (2.516t_6+11.6)\times10^{-4} & 10.37 < t_6 < 37.53 \\
        1.06\times10^{-2} & t_6 > 37.53 \\
    \end{cases}
    \label{fire_rate}
\end{equation}
$N_{\rm SN}$ is the cumulative number of SNe that a stellar population
produces, divided by the initial mass of that population.
The total energy released by a population of stars will simply be the product of
the energy released per SN, $e_{\rm SN}$ and the number of SNe
$E_{\rm SN}= e_{\rm SN} N_{\rm SN}$.  This can then be normalized by the
population mass $M_*$, giving a total specific SN energy $\mathcal{E}_{\rm SN} =
E_{\rm SN}/M_*$.

It is also a common approach to simply detonate all SNe simultaneously, either
at the time of the first SN $t_0\sim3\Myr$, roughly the median SN time
$t_0\sim15\Myr$, or at the time of the last SN $t_0\sim30\Myr$.  As the total
SN number (energy) budget is nearly linear in time, it is also a common
approach to inject SN energy at a constant rate.  A brief sampling of total
specific SN energies $\mathcal{E}_{\rm SN}$, SN start times $t_0$, and
durations $\tau_{\rm SN}$ are given in Table~\ref{other_SN_rates}.  This is in
no way a comprehensive survey of every input rate that has been used to date,
but is an example of what is commonly used, and what is used by recent and
frequently cited simulations.  Many of these models are also used in multiple
works -- for example, the EAGLE feedback models are also used in APOSTLE
\citep{Sawala2016} and E-MOSAICS \citep{Pfeffer2018,Kruijssen2019a} simulations.
As these numbers show, there is both a wide variety in the energy budget for SNe
as well as the timescales over which this energy is deposited.  Instantaneous
and continuous injection of energy are common approaches, while some simulations
use fits to the IMF and the stellar lifetime function to produce more accurate,
if more complex, SN injection rate functions.

\section{Methods}
\label{methods}
We run the simulations presented here in the moving-mesh semi-Lagrangian code
{\sc AREPO} \citep{Springel2010}.  {\sc AREPO} uses an on-the-fly Voronoi
tessellation to generate a mesh which follows the flow of fluid in a simulation
and allows the use of optimally diffusive Riemann solvers in a Godunov-type
scheme, while preserving Galilean invariance and allowing for automatic
Lagrangian refinement.  {\sc AREPO} has been widely used to study galaxy
formation in both large-volume cosmological simulations
\citep{Vogelsberger2014b}, cosmological zooms of individual galaxies
\citep{Grand2017}, down to isolated galaxy simulations \citep{Smith2018}.  We
have added to {\sc AREPO} the {\sc grackle} 3.1 \citep{grackle} non-equilibrium
cooling library, which allows us to include primordial \& metal line cooling
using tabulated {\sc CLOUDY} \citep{Ferland2013} rates.  In this study, we
assume collisional ionization equilibrium and an initial ISM metallicity of $Z=0.012$.
We also include a non-thermal pressure floor to ensure that the
\citet{Truelove1997} criterion is fulfilled and that artificial numerical
fragmentation is suppressed when the Jeans length falls below the cell size.

Star formation is handled in this study using a simple \citet{Schmidt1959} law prescription \citep{Katz1992}.
Stars are allowed to form in gas which has density exceeding $10\hcc$, and with
temperature below $10^4\K$.  Gas cells which satisfy this criterion then form
stars stochastically, with a probability given by the SFR
\begin{equation}
\dot\rho_*=\epsilon_{\rm ff}\rho/t_{\rm ff}
\end{equation}
In a given timestep, the probability of a
gas cell forming a star is thus  
\begin{equation}
P_{\rm SF}(\delta t) =
1-\exp{\frac{-\epsilon_{\rm ff}\delta t}{t_{\rm ff}}}
\end{equation}
We use an efficiency per free fall time of $\epsilon_{\rm ff}=0.05$, appropriate
within molecular clouds \citep[e.g.][]{Evans2009} .  After this probability is
calculated, we draw a random number from a uniform distribution $[0,1)$.  If the
probability $P_{\rm SF}$ exceeds this random number, the cell is fully converted
to a star particle.  As we use a Lagrangian refinement that ensures gas cells
never vary by more than a factor of 2 from our target mass resolution of
$8.59\times10^5\Msun$, this also ensures the star particles spawned in our
simulation stay close to this target mass as well. While
there are more complex models for star formation, the interplay of these models
and stellar feedback is beyond the scope of this study.

As simply dumping thermal energy into gas will result in catastrophic
overcooling when the Sedov radius of an individual SN is unresolved, we
make use of a mechanical feedback model that has been demonstrated to produce
converged momentum injection across many orders of magnitude in gas resolution
\citep{Kimm2014,Hopkins2018a}.  In brief, we determine the host cell of a star
particle depositing feedback, and then use the Voronoi mesh around this cell to
deposit momentum in the shell of the feedback bubble.  For each cell which
shares a face with the feedback-hosting cell, we calculate the area of that face
$w_i = A_{ij}$, and use that to weight the momentum contribution for each cell.
As {\sc AREPO} provides both face areas and normals, we are able to ensure
momentum conservation trivially.  We use the terminal momentum at the end of the
pressure-driven snowplow phase, following \citet{Kimm2014}, using the equations
derived from high resolution one- and two-dimensional simulations of SN blasts:
\begin{equation}
    p_{\rm term}  =
    3\times10^5\kms\Msun\max{[Z/Z_\odot,0.01]}^{-0.14}E_{51}^{16/17}n^{-2/17}
    \label{blondin_momentum}
\end{equation}
For each cell surrounding the central one, we use the energy $E_i=w_iE$ and density
to calculate $p_{\rm term}$, and inject the minimum of the ejecta momentum or the
terminal momentum:
\begin{equation}
    p_{ij} = \min{[\sqrt{2m_jw_iE}, w_ip_{\rm term}]}
    \label{mechanical_momentum}
\end{equation}
This is essentially the same scheme used in \citet{Hopkins2018a} and \citet{Smith2018}, but
with one significant difference.  In those methods, the remaining feedback
energy after momentum injection is deposited into the same cells which receive
momentum.  We instead collect this thermal energy and deposit it into the central
cell, effectively giving us a separation between the cold, swept up shell (which
carries most of the momentum), and the hot, diffuse centre (which contains most
of the thermal energy).  We have determined that this results in little
difference in the overall evolution of SN driven bubbles in isolation, as well
as on the overall evolution of galaxies simulated with this feedback model
compared to simply injecting both thermal and kinetic energy into the same cells.
What it provides is the ability for more sophisticated treatments of
marginally-resolved SN bubbles, which we will present in a future study.

We use the initial conditions (ICs) developed as part of the AGORA comparison
project \citep{Kim2014}.  This IC was designed to roughly match the Milky Way:
it has a disc scale length of $3.43\kpc$, and a scale height a tenth of this value.
It is embedded in a dark matter (DM) halo with a mass $M_{200}=1.07\times10^{12}\Msun$ and
virial radius $R_{200}=205\kpc$.  The halo concentration parameter is $c=10$,
and the \citet{Bullock2001} spin parameter is $\lambda=0.04$.  The IC contains a
stellar disc and bulge, with a bulge to stellar disc ratio of $0.125$, and a gas
fraction of $0.18$.  The AGORA isolated disc ICs were generated
using the {\sc MAKENEWDISK} code described in \citet{Springel2005}.  We use a
gravitational softening length of $80\pc$, and a gas cell mass of
$8.59\times10^4\Msun$.    The IC star particle mass is $3.437\times10^5\Msun$,
and the live DM halo contains $10^5$ particles of mass
$1.254\times10^7\Msun$ each.  We initialize the gas in the simulation with a
temperature of $10^4\K$, though this is rapidly replaced with the equilibrium
temperature determined by the ISM density and the \citet{Haardt2012} UV
background.

\section{Does it matter when supernovae occur?}
In this section, we examine how the evolution of an isolated disc galaxy
changes when the timing of SNe is changed while keeping the total SN energy budget
constant.  We will examine both instantaneous injection, where $\tau_{\rm
SN}=0$, and continuous injection, where $\tau_{\rm SN} > 0$.  In each case, the
total mass loss due to SNe is set to $10$ per cent of the star particle's initial mass,
and the energy budget is set by this initial mass with the specific SN energy of
$\mathcal{E}_{\rm SN}=10^{49} \erg/\Msun$, corresponding to an energy released
per SN of $10^{51}\erg$ with $N_{\rm SN}=0.01\Msun^{-1}$.

\subsection{Instantaneous injection and the SN delay time}
\label{delay_time}
\begin{figure}
    \includegraphics[width=0.5\textwidth]{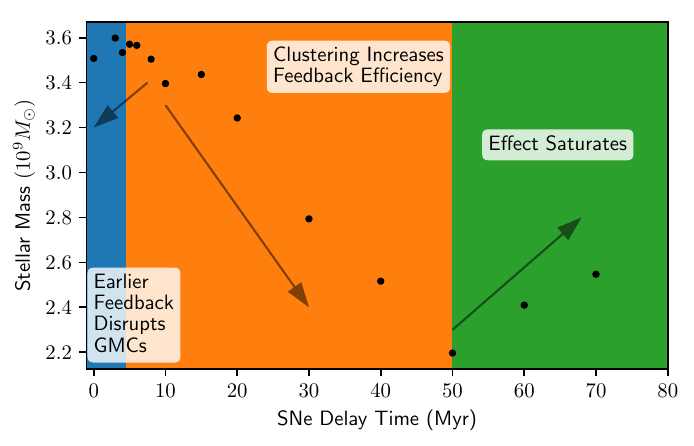}
    \caption{Stellar mass formed over $600\Myr$ as a function of SN delay time,
    shown as black points.  We can see three regions (indicated here with
    different colors) where the relation between the total stellar mass and the
    delay time changes.  We indicate three different regimes here.  In blue,
    shorter delays reduce the SFR by disrupting star forming clouds before they
    reach high star formation efficiencies.  In orange, we see the regime where
    increased clustering increases the efficiency of SN feedback.  Finally, in
    the green region, we see this effect saturate, as stars formed co-spatially
    drift apart.}
    \label{stellarmass_delay}
\end{figure}
\begin{figure}
    \includegraphics[width=0.5\textwidth]{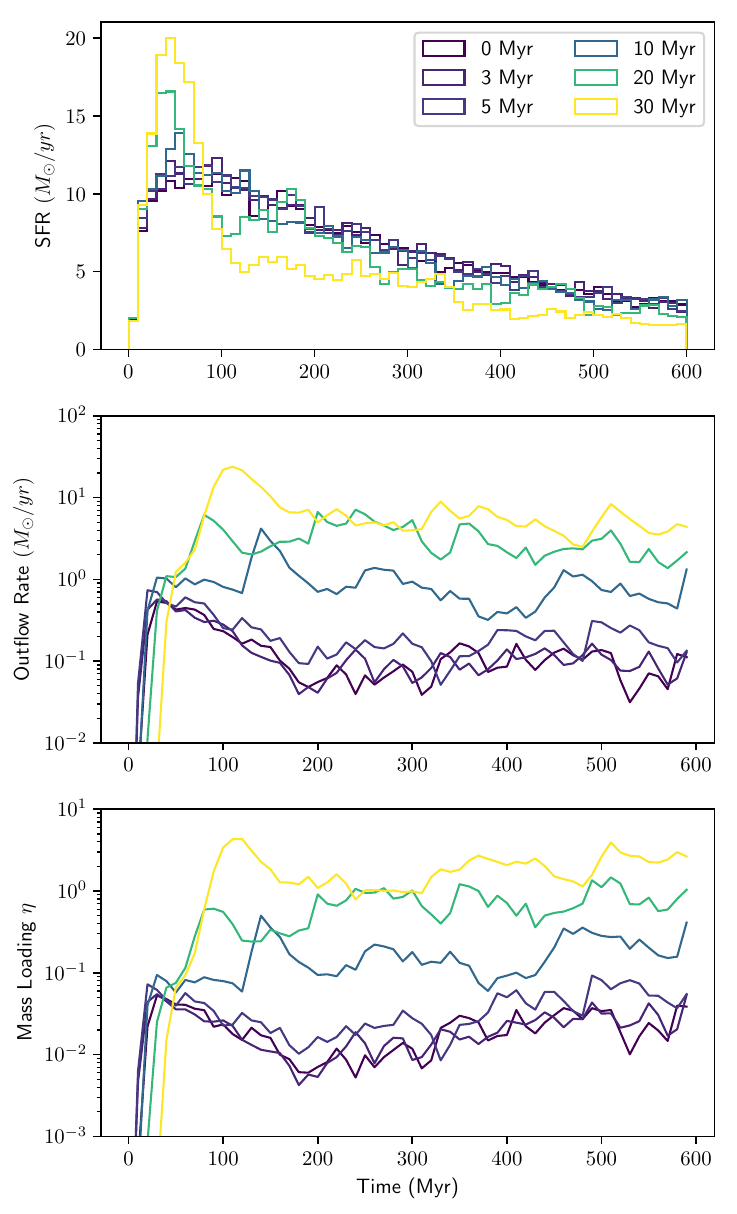}
    \caption{Star formation rates (top panel), outflow rates (centre panel), and
    mass loadings (bottom panel) for different SN delay times.  The top panel
    shows that long SN delays ($t_0=20-30\Myr$) have clearly lower overall  star
    formation rates, and that the intensity of the initial burst is larger for
    longer delays.  As the middle and bottom panels show, the difference in
    outflow rates and mass loadings for $t_0=0-5\Myr$ is negligible, while
    longer delay times result in significantly higher outflow rates and mass
    loadings, even long after the initial starburst has abated.  Going from a
    delay of $5\Myr$ to $30\Myr$ reduces the average star formation rate for the
    final $400\Myr$ of the simulations by approximately one order of magnitude,
    but increases the mass loading by over two orders of magnitude.}
    \label{sfr}
\end{figure}
\begin{figure*}
    \includegraphics[width=\textwidth]{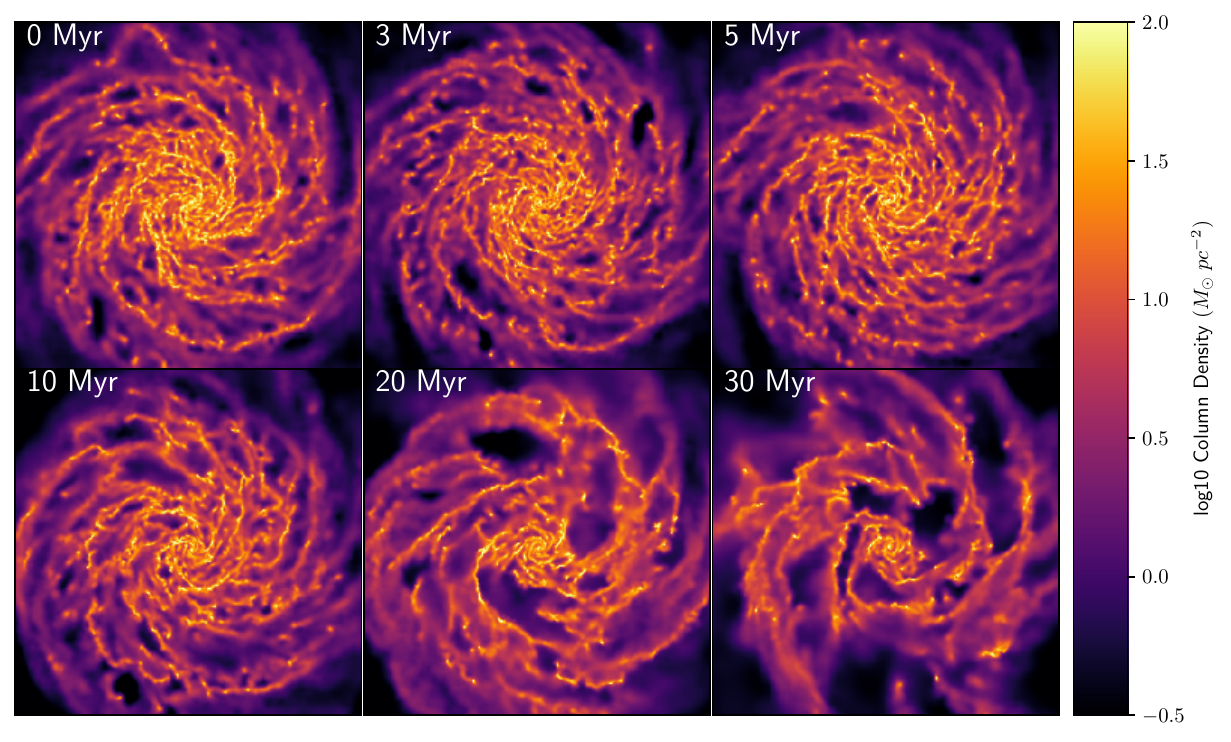}
    \caption{$30\kpc$ wide gas column density maps for different SN delay
    times.  As can be seen here, longer delay times result in a much ``emptier''
    ISM, with efficient outflows removing a large fraction of the gas from the
    disc.  Much of the remaining mass is concentrated in a small number of dense
    clouds.}
    \label{gas_column}
\end{figure*}
\begin{figure}
    \includegraphics[width=0.5\textwidth]{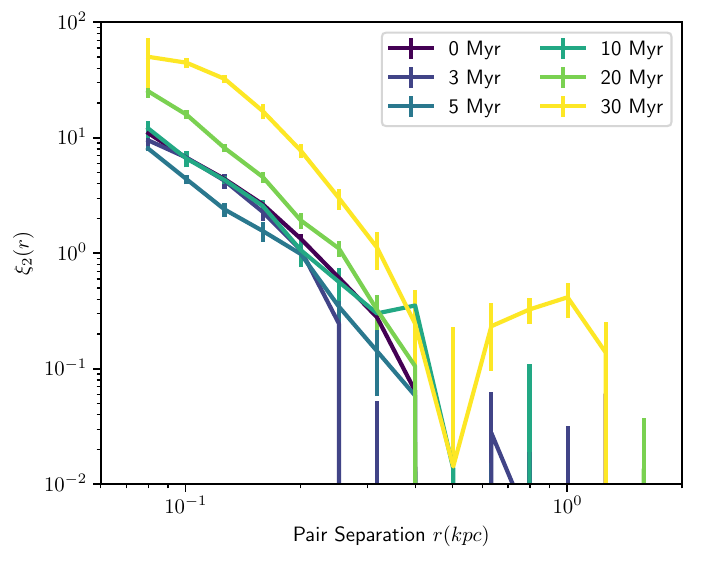}
    \caption{Two point correlation function $\xi_2(r)$ for young star particles
    (with ages $<100\Myr$) in our simulations as a function of the SN delay time
    $t_0$.  For long delay times ($t_0>10\Myr$), clustering on $<200\pc$ length
    scales becomes significantly more likely than in the galaxies with shorter
    delay times.}
    \label{2point}
\end{figure}
\begin{figure}
    \includegraphics[width=0.5\textwidth]{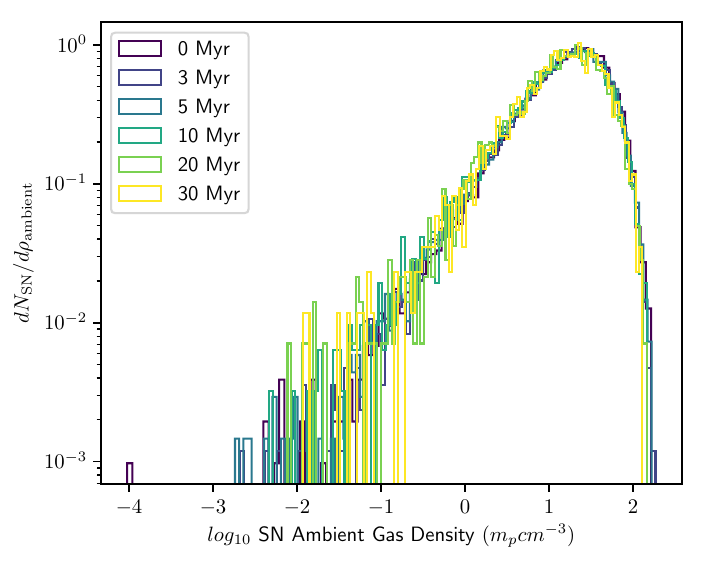}
    \caption{Local ISM density of SN events, as a function of the SN delay time.
    As can be seen here, there is no significant change in the ambient density
    of the ISM where SNe detonate when the SN delay time is changed.  This
    implies that the increased feedback efficiency is not due to stars migrating
    out of dense regions before SNe, or gas being consumed by star formation
    before SNe occur.}
    \label{sn_rho}
\end{figure}
Here we examine the results of dumping the entire SN energy budget
instantaneously, after some delay time.  As we previously showed in
Table~\ref{other_SN_rates}, instantaneous SN injection is frequently used in
many simulations that include SN feedback. In Figure~\ref{stellarmass_delay}, we
show the total stellar mass formed over the $600\Myr$ runtime of our simulations
as a function of the delay time $t_0$.  As can be seen, there is a significant,
nonlinear effect of the SN delay time on the averaged SFR/total stellar mass
formed. For very short delays $<5\Myr$ (shown in blue), shorter delays can
slightly reduce the star formation by disrupting star forming clouds before they
have a chance to fully collapse.  For longer delay times, $\sim5-50\Myr$, the
star formation efficiency of individual clouds increases, resulting in much more
clustered star formation, which subsequently drive stronger outflows.    With
delay times of $t_0=5-50\Myr$, there is significantly lower star formation with
longer delay times, with a nearly $50$\% drop in stellar mass produced with
$t_0=50\Myr$ compared to $t_0=5\Myr$.  Delays beyond this ($>50\Myr$, shown in
green) reverse the trend, as the local star formation efficiency in clouds has
saturated at $\sim1$, and stars which were once more clustered drift apart. In
order to verify the robustness of these results to run-to-run stochasticity, we
re-simulate three of these delay times in Appendix~\ref{stoch}, and show that
the differences in SFR, outflow rates, and mass loadings are greater than
run-to-run variance.

We show in Figure~\ref{sfr} how varying $t_0$ affects the evolution of three
important quantities that are determined by the effectiveness SN feedback.  In
the top panel, we show the star formation rate, which shows how SN feedback is
able to slow and regulate star formation.  In the middle panel, we show the gas
outflow rate ($\dot M_{out}$) from the galaxy, which shows how well SN feedback can drive
galactic winds and fountains.  Finally, we show in the bottom panel the mass
loading ($\eta = \dot M_{\rm out} / {\rm SFR}$), the ratio of the two above
quantities, which allows us to distinguish between outflows driven by efficient
SN feedback from those driven by inefficient feedback, but high SFRs.  

The reduced SFRs with long ($t_0 \geq 10\Myr$) delays is not uniform across the
$600\Myr$ evolution of the galaxy.  The top panel of Figure~\ref{sfr} shows that
long delays produce a strong burst of star formation at the beginning of the
simulation, as clouds are able to form and rapidly form stars without any
regulation from SN feedback.  This leads to a significant burst in star
formation in the first $100\Myr$, followed by a significant reduction in the
overall SFR for the remaining time.  We verify, in Appendix~\ref{starburst},
that this burst is not the source of the difference between simulations with
different values of $t_0$.   In the centre panel of the same figure, we
see that this burst in star formation is associated with a corresponding burst
in gas outflowing from the galaxy.  We calculate these outflow rates by taking
all gas moving away from the disc within two planar slabs of thickness $500\pc$
located at $5\kpc$ above and below the disc.  The outflow rate is then
calculated, as in \citet{Keller2014}, as the total momentum of outflowing gas in
these slabs, divided by their thickness.  After the initial starburst, the
outflow rate is higher for longer delay times, with delays of $10-30\Myr$ giving
comparable outflow rates of $\dot M_{\rm out}\sim 1-10\Msun{\rm yr}^{-1}$, and shorter
delays producing outflow rates of $\dot M_{\rm out}\sim0.1\Msun{\rm yr}^{-1}$.
If we look at the mass loading $\eta$, we can see that there is a monotonic
increase in mass loading with increased $t_0$.  This means that longer SN delay
times will result in SN feedback that more effectively drives galactic winds,
{\it without any changes to the SN energetics or numerical coupling algorithm}.
The effect is especially striking when we compare the shortest delay ($0\Myr$)
to the longest ($30\Myr$), with average mass loadings of $\eta=0.021$
and $\eta=1.7$ respectively, an increase of nearly two orders of magnitude.

The increased effectiveness of feedback at driving outflows is qualitatively
striking when we look at gas column density maps for different SN delay times in
Figure~\ref{gas_column}.  The stronger outflows we saw previously for long
delay times begin to significantly alter the disc morphology for delays of
$t_0 \geq10\Myr$.  We see that the ISM becomes noticeably depleted, with
nearly all gas being isolated to dense clumps and clouds along spiral arms, as
the flocculent inter-cloud material is evacuated.  Overall, we see much less
dense ISM gas in the disc of galaxies with longer values of $t_0$.

The reason for the increased effectiveness of SN feedback is the way
that the SN delay time $t_0$ changes the clustering of star formation, which we
can quantify with the stellar two point correlation function.  As we show in
Figure~\ref{2point}, the two point correlation function of stars recently formed
(within the last $100\Myr$) in the disc has significantly more clustered star
particles with separations of $<300\pc$ when we increase $t_0$ beyond $10\Myr$.
We calculate the two point correlation function $\xi_2(r)$ using the
\citet{Landy1993} estimator\footnote{The two point correlation function
$\xi_2(r)$ is a measure of the excess probability of finding two stars within a
separation of $r$ against a random distribution.  The \citet{Landy1993}
estimator uses the number of actual pairs within a separation $r$, $DD(r)$,
together with the number of random pairs given the same mean density $RR(r)$ and
the cross-correlated data-random pairs $DR(r)$ to calculate $\xi_2(r) =
(DD(r)-2DR(r)+RR(r))/RR(r)$.  As we are dealing with star particles distributed
in a thin disc, the \citet{Landy1993} estimator is ideal, as it minimizes the
errors occurring from a non-periodic distribution of points.} with uncertainties
calculated with 10 bootstrap resamples.  In order to exclude any structures
``baked in'' from the initial star burst and the symmetry of the initial
conditions, we only look at stars formed in the final $100\Myr$ of the
simulations.  This also allows us to look specifically at stars that have not
experienced significant migration away from the stars they are born along side.
For delay times of $t_0\geq10\Myr$, we see little difference in $\xi_2(r)$, which
matches the small changes we see in the SFR, galactic outflow rate, and ISM
morphology shown in Figure~\ref{sfr} and Figure~\ref{gas_column} (although the
two point correlation function for $t_0=10\Myr$ shows only marginally greater
clustering than the runs with $t_0<10\Myr$, within the bootstrapped
uncertainties of $\xi_2(r)$).  The much greater probability of
forming stars in clusters with long values of $t_0$ is simply a result of
whether feedback is able to disrupt the star formation of a local region before
simple gas exhaustion or galactic dynamics do so \citep{Kruijssen2012c}.  With
short delay times, star forming regions form only a fraction of the available
gas into stars, while long delay times allow star forming regions to reach their
maximum integrated star formation efficiencies. A possible alternative
explanation for the increased feedback efficiency with long delay times may be
the ISM density in which SNe detonate.  At low density, radiative cooling losses
are reduced, and so feedback may couple more efficiently with the ISM.  If long
delay times allow more star forming gas to be consumed, or for stars to migrate
out of their birth environments, they may detonate as SNe in significantly lower
density gas.  However, as Figure~\ref{sn_rho} shows, there is no significant
change in the density of gas in which SNe detonate as a function of $t_0$.  If we
restrict this to only examine stars formed in the last $100\Myr$, as we do in
the correlation functions, the distribution shows little change. This strongly
suggests that it is the clustering, not the ambient ISM density, that increases
the efficiency of instantaneously injected feedback with long delay times.

\subsection{Instantaneous vs. continuous injection of energy}
\label{continuous}
\begin{figure}
    \includegraphics[width=0.5\textwidth]{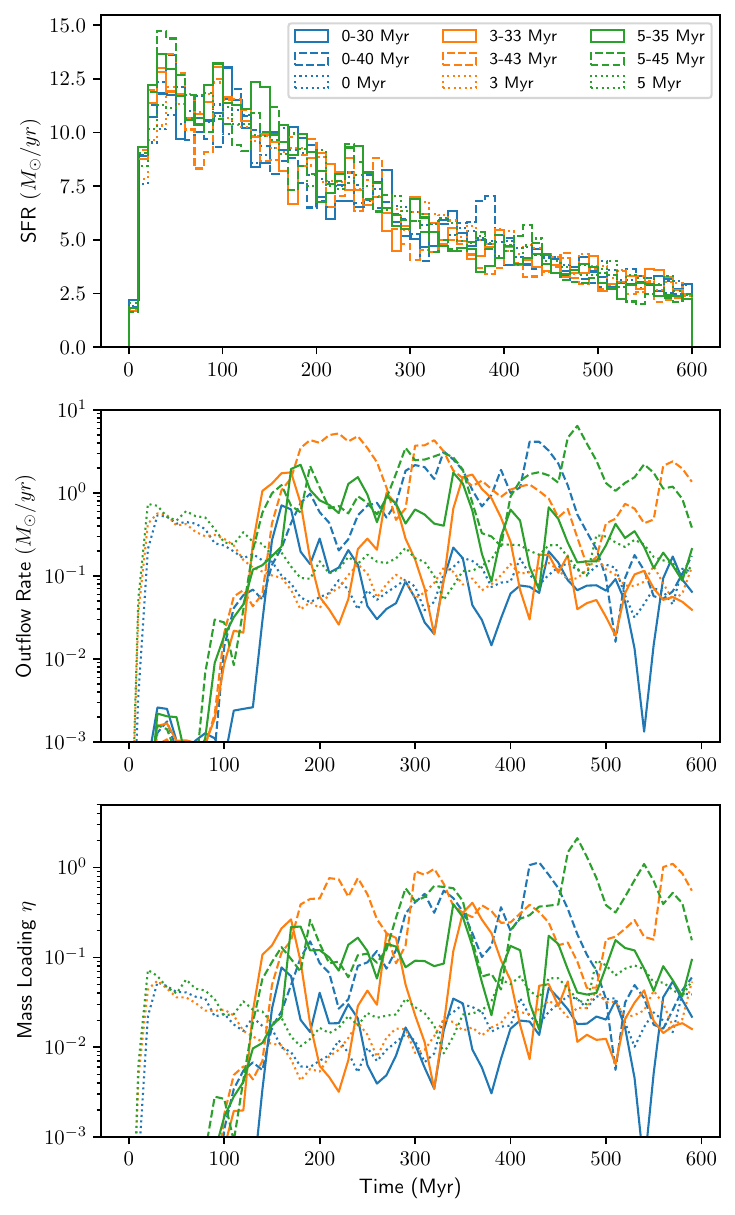}
    \caption{Star formation rates (top panel), outflow rates (centre panel), and
    mass loadings (bottom panel) for continuous SN injection with different
    delay times $t_0$ and durations $\tau_{\rm SN}$.  Solid lines show results using
    a duration of $\tau_{\rm SN}=30\Myr$, while dashed lines show
    $\tau_{\rm SN}=40\Myr$.  Dotted lines show the results for the same delay times
    with instantaneous injection. Blue, orange, and green curves show delay
    times of $0\Myr$, $3\Myr$, and $5\Myr$ respectively.  Longer delay times
    result in a slightly lower SFR, with little discernible
    difference in outflow rates.}
    \label{sfr_continuous}
\end{figure}
\begin{figure}
    \includegraphics[width=0.5\textwidth]{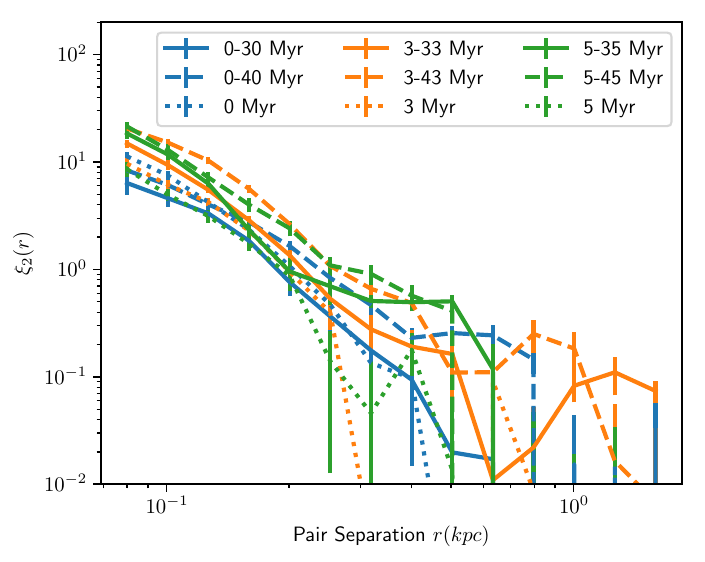}
    \caption{Two point correlation function $\xi_2(r)$ for star particles in our
    simulations as a function of the SN delay time $t_0$ and injection
    timescale $\tau_{\rm SN}$.  As in Figure~\ref{sfr_continuous}, solid lines show
    $\tau_{\rm SN}=30\Myr$, dashed lines show $\tau_{\rm SN}=40\Myr$, while dotted lines
    show instantaneous injection with the same delay time $t_0$.}
    \label{2point_continuous}
\end{figure}
\begin{figure}
    \includegraphics[width=0.5\textwidth]{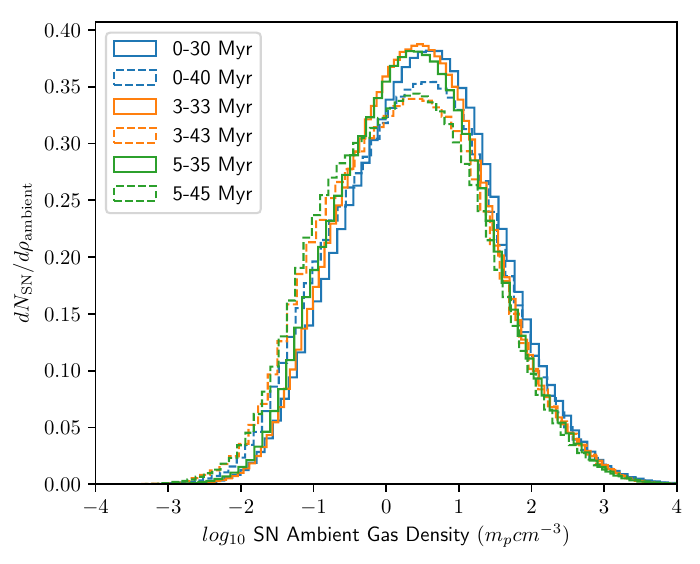}
    \caption{Local ISM density of SN events with continuous injection of SN
    energy (for comparison to instantaneous injection, see Figure~\ref{sn_rho}).
    As can be seen, the typical densities of SN detonations during continuous
    injection are lower than those seen with instantaneous injection, with
    roughly log-normal distribution of ambient ISM densities as well.  With
    longer ($40\Myr$) injection durations, we see a somewhat broader
    distribution of ambient densities, with a lower median density as well.}
    \label{sn_rho_continuous}
\end{figure}
While the instantaneous injection of SN energy is a common technique in galaxy
simulations, it is of course less physically motivated than using a realistic
distribution of stellar initial masses and lifetimes, which together can produce
an SN injection rate, as we derived in Section~\ref{injection_rates}.  We can
reasonably approximate the injection rate for a Salpeter-like IMF and stellar
lifetime function as a piecewise constant luminosity if the IMF in each stellar
particle is fully sampled, and the number of SNe produced per star particle is
more than $\sim10$ \citep{MacLow1988}.  As our stellar particle mass is
$\sim10^5\Msun$, this approximation holds comfortably.  With a continuous
injection of energy, we add a second timing parameter to the SN feedback, the
duration of energy injection $\tau_{\rm SN}$.  Here we probe within the
range of ``reasonable'' physical uncertainty for $t_0$ and $\tau_{\rm SN}$.  We simulate
a grid of 6 galaxies, with delay times $t_0$ of $0\Myr$, $3\Myr$, and $5\Myr$ to
approximate ``early feedback'' (as is done in \citealt{Semenov2018}), rapid SN
onset, and slow SN onset respectively.  We choose two durations times, $30\Myr$ and
$40\Myr$, that span the range of durations that have been commonly used in the
literature.

Figure~\ref{sfr_continuous} shows that for similar delay times to instantaneous
injection, continuous injection of SN energy produces roughly comparable star
formation rates and higher, burstier outflow rates and mass loadings, once the
disk settles to a steady state.  As with instantaneous injection, there is
little difference in either star formation or outflow properties for changes in
delay times of $0-5\Myr$, and we see that there is no noticeable change in the
SFR for SN durations of $40\Myr$ as opposed to  $30\Myr$.  Generally, there are
higher outflow rates and mass loadings for runs with longer durations
($\tau_{SN}=0$ has lower average mass loadings for all runs with $\tau_{SN}>0$,
with the exception of the $0-30\Myr$ case).  The average mass loadings for the
three values of $t_0$ we examine here are $\eta=0.026$ with $\tau_{SN}=0$,
$\eta=0.05$ with $\tau_{SN}=30\Myr$, and $\eta=0.25$ for $\tau_{SN}=40\Myr$.
However, the galactic outflows in these galaxies are delayed in their launching
from the initial starburst, peaking at $100-200\Myr$, rather than in the first
$50\Myr$ when $\tau_{SN}=0$.  Interestingly, if we look at the same two point
correlation function of star particles as we did for the instantaneous case,
seen here in Figure~\ref{2point_continuous}, we see that the continuous
injection of SN energy does not significantly change clustering on scales
$<100\pc$ compared to instantaneous injection (though we do see that the case
with the lowest outflow mass loading, $0-30\Myr$ also shows the lowest
clustering of star formation).  Continuous energy injection does, however,
produce noticeable changes in the ambient density that SNe explode in.  As
Figure~\ref{sn_rho_continuous} shows, the ambient density surrounding SN events
is approximately log-normal, with lower mean densities than are seen for
instantaneous injection.  With instantaneous injection of SN energy, we see
ambient ISM densities about SN events of $15-17\hcc$.  With continuous injection,
longer durations produce lower ambient densities. For $\tau_{\rm SN}=30\Myr$ we see
ambient densities of  $3.2\hcc$, $2.7\hcc$, and $2.4\hcc$ with $t_0=0,3,5\Myr$
respectively.  For $\tau_{\rm SN}=40\Myr$, we see ambient densities of  $2.4\hcc$,
$1.9\hcc$, and $1.7\hcc$ with $t_0=0,3,5\Myr$ respectively.

\begin{figure*}
    \includegraphics[width=\textwidth]{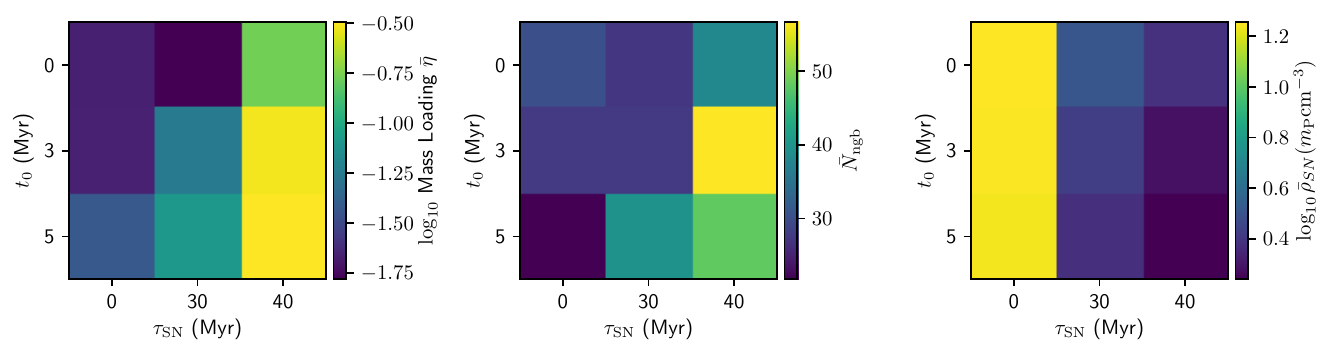}
    \caption{Summary of the average outflow mass loading (left panel), average
    number of young stellar neighbours per star particle (middle panel), and
    ambient SN density (right panel) for simulations with SN delay times of
    $t_0=0-5\Myr$ and injection durations $\tau_{\rm SN}=0-40\Myr$.  As can be
    seen here, mass loadings increase towards larger values of $t_0$ and
    $\tau_{\rm SN}$.  A similar trend is seen in the ambient gas density
    surrounding SN events (though this trend is stronger for continuous SN
    energy injection.  We also see a rough trend towards more clustered young
    star particles following this same trend.)}
    \label{matrix}
\end{figure*}

Given these results, we naturally might ask why the significantly lower ambient
SN densities seen for SN injected continously (a factor of 2-5 times lower than
for SN injected instantaneously) do not result in significantly higher outflow
mass loadings?  If the primary physical mechanism which sets the efficiency of
outflow driving is radiative cooling in the ISM, we should see much higher mass
loadings for the simulations which show lower ambient SN gas density.
As we show in the summary Figure~\ref{matrix}, the highest averaged mass
loadings for SN injection with $t_0=0-5\Myr$ and $\tau_{\rm SN}=0-40\Myr$  do
correspond to the lowest average densities.  However, we do not see a sharp
change in mass loadings going from instantaneous to continuous injection of SN
energy that corresponds to the sharp change in ambient SN gas densities shown in
the right hand panel of Figure~\ref{matrix}.    If the decrease in ambient
density was combined with a corresponding decrease in stellar clustering, this
might explain the relatively weak increase in outflow mass loadings we see. We
also do not see a sharp change in the average number of young ($<100\Myr$) stars
within one scale height ($343\pc$) of every star particle as we move from
instantaneous to continuous star formation.  A different mechanism than the
clustering of young stars is thus needed to explain the relatively small
increase mass loadings we see for continous injection of energy relative to
instantaneous injection.  

If we consider the evolution of a pressure-driven snowplow (for instantaneous
injection) versus a wind (for continuous injection), we can see a possible
solution to this question.  Outflows will vent from the ISM once a superbubble
radius reaches a scale height $r\sim h$.  In both cases, a swept-up
shell will rapidly form, with radiative cooling removing most of the thermal
energy from this shell.  For instantaneous injection, this shell will evolve
with the familiar pressure-driven snowplow solution derived by \cite{McKee1977}.
In this case, the radius of the bubble (from equation 9 and 12 in \citealt{McKee1977}) will
be given by the driving energy ($E_{51}$, in units of $10^{51}\erg$), the
ambient density ($n_0$, in units of $\hcc$), and the time ($t$, in units of
years):
\begin{equation}
    r = 10^{0.33}E_{51}^{0.15}\left(\frac{t^2}{n_0}\right)^{1/7} \pc .
    \label{pds_r}
\end{equation}
Setting this $r=h$ (in $\pc$) to solve for the breakout time gives:
\begin{equation}
    t_{\rm b,inst} = 10^{-7.16}E_{51}^{-0.53}n_0^{1/2}h_{\pc}^{7/2} \Myr.
\end{equation}
We can then differentiate equation~(\ref{pds_r}) and substitute $t_{\rm b,
inst}$ for $t$ to solve for the breakout velocity:
\begin{equation}
    v_{\rm b,inst} = 10^{6.82}E_{51}^{0.53}n_0^{-1/2}h_{\pc}^{-5/2} \kms .
\end{equation}
For a star cluster driving an energy conserving wind we can apply the classic
\citet{Weaver1977} solution for the bubble radius:
\begin{equation}
    r = 10^{-1.77}L_{38}^{1/5}t^{3/5}n_0^{-1/5} \pc
    \label{weaver_r}
\end{equation}
with a mechanical luminosity $L_{38}=(E/\tau_{\rm SN})/10^{38}\ergs$.  
To solve for the shell velocity, we first set $r_{\rm b} = h$ (in pc) and invert
equation~(\ref{weaver_r}) to solve for $t$:
\begin{equation}
    t_{\rm b,cont} = 10^{-3.04}L_{38}^{-1/3}n_0^{1/3}h_{\pc}^{5/3} \Myr .
\end{equation}
By then differentiating equation~(\ref{weaver_r}) and substituting in $t=t_{\rm
b,cont}$ we find the velocity of a continuously-driven blastwave at the time of
breakout is 
\begin{equation}
    v_{\rm b,cont} = 10^{2.81}L_{38}^{1/3}n_0^{-1/3}h_{\pc}^{2/3} \kms.
\end{equation}
For a single star particle with mass $8.9\times10^4\Msun$ (and thus
$E=8.9\times10^{53}\erg$), in an ISM with ambient density $15\hcc$ and a scale
height of $343\pc$, the velocity of a bubble at breakout for instantaneous
injection of energy is $v_{\rm b,inst}=28\kms$, which occurs at $t_{\rm b,
inst}=5.5\Myr$.  For the same star particle, injecting its SN energy over
$\tau_{\rm SN}=30\Myr$ into an ambient density of $2\hcc$, this breakout is
instead $v_{\rm b,cont}=22\kms$, and occurs at $t_{\rm b, cont}=9.1\Myr$.  If
the mass loading of the galactic winds are primarily set by the velocity at
breakout, this explains why the larger ambient densities seen for instantaneous
injection of SN energy does not result in a significant reduction of the outflow
mass loadings compared to the continuous injection of SN energy. Naturally, this
one-dimensional analysis omits the stratification of gas within the galaxy, as
bubbles will become hourglass-shaped as their radius becomes comparable to the
gas scale height \citep{MacLow1988}, but is illustrative of how bubbles driven
by an instantaneous blastwave evolve relative to those driven by a constant
mechanical luminosity.

\begin{figure}
    \includegraphics[width=0.5\textwidth]{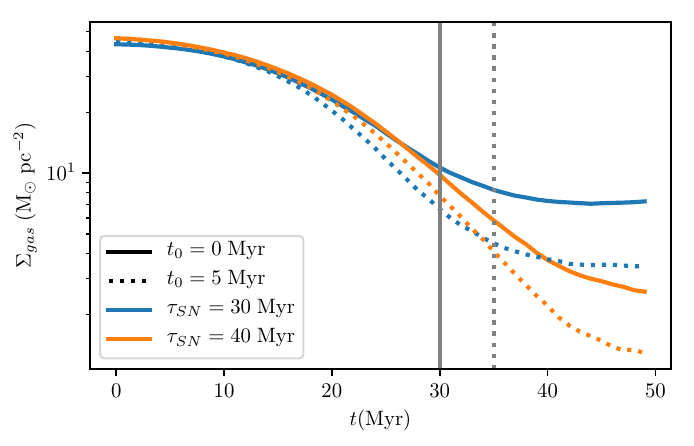}
    \caption{Gas surface density within a scale height of newly-formed star particles
    for their first $50\Myr$ with continuous injection of SN.  We show the lower
    25th percentile surface density to highlight the quarter of stars which are
    most able to drive outflows efficiently.  Blue curves show the surface
    density evolution for $\tau_{SN}=30\Myr$, while orange curves show the
    results for $\tau_{SN}=40\Myr$.  Solid curves show $t_0=0\Myr$, and solid
    curves show $t_0=5\Myr$.  The horizontal grey lines indicate the shut off
    point for $\tau_{SN}=30\Myr$.  Stars
    continuously reduce the gas surface density in their neighbourhood until the
    termination of feedback.  With longer injection durations, this results in
    lower final surface densities $50\Myr$ after a star particle has formed.
    Only after the shut off of SN do the $\tau_{SN}=30\Myr$ and
    $\tau_{SN}=40\Myr$ cases begin to diverge significantly.}
    \label{sigma_continuous}
\end{figure}

Despite the small differences in the average local density of SN injection and
the clustering of stars for different values of $\tau_{SN}$, we see higher
outflow rates for galaxies simulated with longer ($\tau_{SN}=40\Myr$) compared
to shorter ($\tau_{SN}=30\Myr$) injection durations.  The previous analytic
argument predicts that bubbles should break out of the ISM before the end of
$\tau_{SN}$.  If this is the case, then SN occurring after the breakout occurs
will be able to accelerate gas out of the disc unimpeded by the weight of the
ISM above or below them.  In Figure~\ref{sigma_continuous}, we examine how the
gas surface density surrounding young stars is altered by the injection of SN.
This figure shows the evolution of the gas surface density within a circular
patch with radius equal to the initial gas scale height $r=h=343\pc$ around each
star particle for $50\Myr$ after formation.  To probe the
stars most likely to launch strong outflows, we calculate the lower 25th
percentile of these profiles.  The surface densities for all cases begin
at $\sim40\Msun\pc^{-2}$, and are reduced to $<10\Msun\pc^{-2}$ by $50\Myr$.  As
can be seen, however, more gas is removed from the regions surrounding stars
when $\tau_{SN}=40\Myr$ than when $\tau_{SN}=30\Myr$, with surface densities
$\sim4$ times lower by $50\Myr$. This difference does not begin until the
termination of SN in the $\tau_{SN}=30\Myr$ case, suggesting that the size of
SN-driven bubbles (and/or the ejection of outflows) depends weakly on $\tau_{SN}$ prior to the shut off of SN (as the
relatively weak dependence on $L_{38}$ in equation~\ref{weaver_r} would
predict).  If these bubbles do indeed break out before $\tau_{SN}$, then the
energy and momentum injected by SN will be able to vent directly into the halo.
A larger fraction of the SN energy budget will be expended in this regime when
$\tau_{SN}$ is longer, resulting in larger outflow mass loadings.  This
phenomenon is analogous to the ``Powered Break-out'' versus ``Coasting
Break-out'' regime that was recently explored in \citep{Orr2021}, with longer
values of $\tau_{SN}$ spending longer in the ``Powered Break-out'' regime.
Further study (such as high resolution simulations of instantaneous and
continuous energy injection in a turbulent, fractal ISM) can potentially explain
the nonlinear connection between the injection timescales of SN, radiative
cooling losses, correlated star formation, and the breakout of superbubbles to
drive galaxy-scale outflows.  

In the next section, we show how the longer durations driven by changes in the
minimum SN progenitor mass have an even stronger effect on the ambient density
in which SNe detonate. This in turn drives even larger changes in the outflow
rates than those we see here, due also in part to changes in the total SN energy
budget. We also do not see any trend in the smallest-scale ($<100\pc$)
clustering between the two feedback durations we simulate, though in the
intermediate ($100-400\pc$) scale, longer delays do appear to produce slightly
more power in the two point correlation function.  We leave a detailed study of
how feedback shapes the clustering of star formation on the smallest (unresolved
in these simulations) scales for future, higher resolution simulation work.

\section{The critical importance of the minimum progenitor mass}
\label{SN_min_mass}

\begin{figure}
    \includegraphics[width=0.5\textwidth]{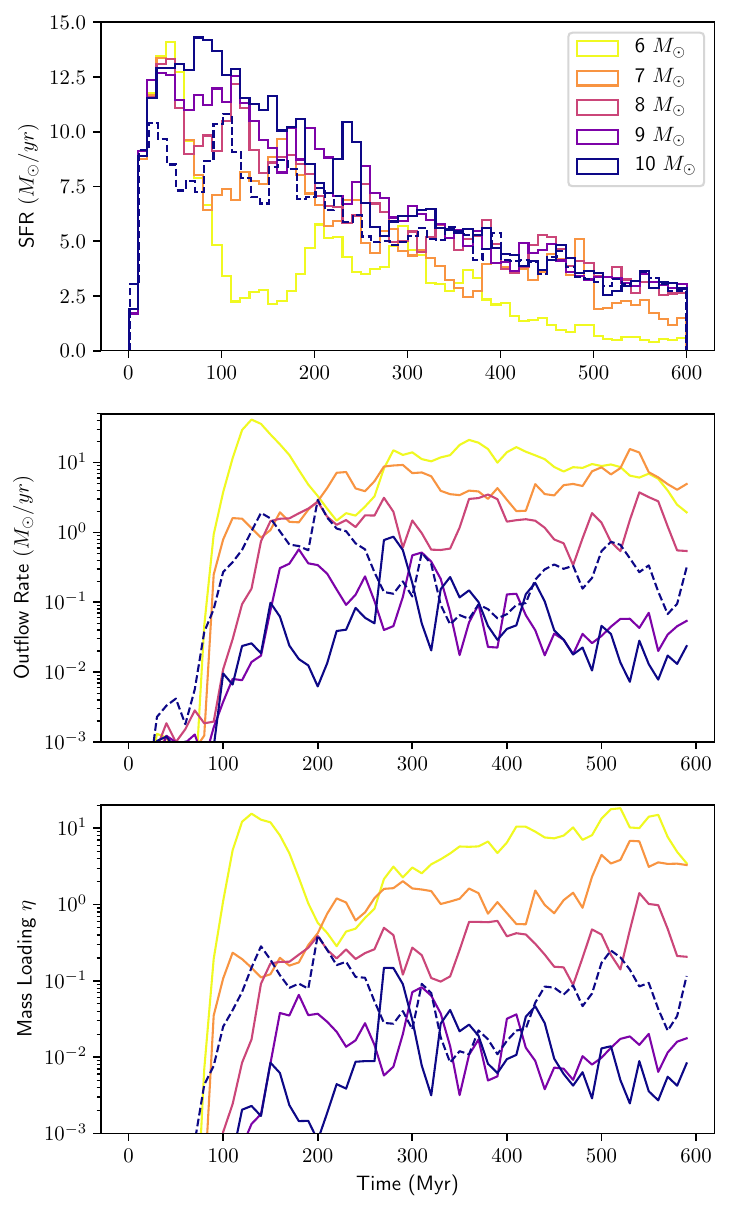}
    \caption{Star formation rates (top panel), outflow rates (centre panel), and
    mass loadings (bottom panel) for continuous SN injection with different
    minimum SN progenitor masses.  All solid curves show the results with the
    feedback energy and duration for a given minimum SN progenitor mass (see
    table~\ref{SN_min_mass_inputs}).  The single dashed curve shows the a
    simulation with the energy budget of a $6\Msun$ minimum progenitor mass, but
    with the shorter duration of a $10\Msun$ minimum progenitor mass.
    As the top panel shows, the larger SN budget
    and longer duration of SN feedback greatly reduces the overall star
    formation rate when lower minimum SN progenitor masses are used.  This
    results in a corresponding increase in the outflow mass loading, as fewer
    stars are able to drive comparable amounts of material out of the galactic
    disc.}
    \label{sfr_minmass}
\end{figure}
\begin{figure}
    \includegraphics[width=0.5\textwidth]{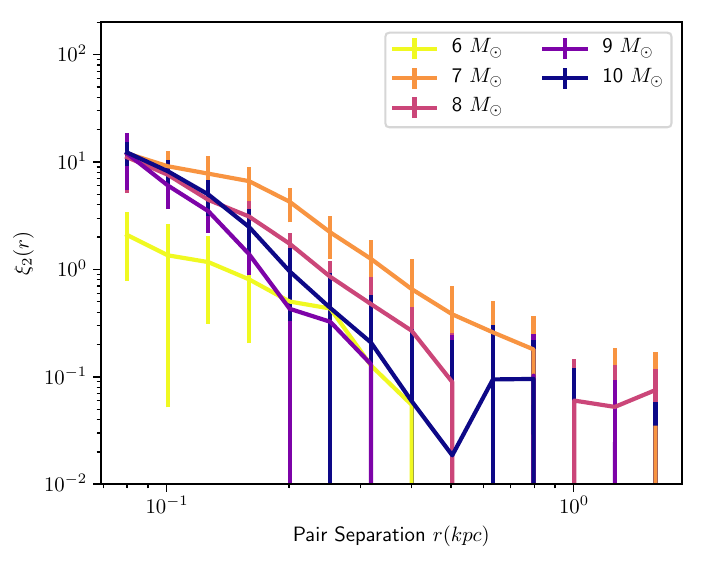}
    \caption{Two point correlation function $\xi_2(r)$ for star particles in our
    simulations as a function of the minimum SN progenitor mass.  Unlike the
    cases with longer SN delay times, we see here that the reduced SFR and
    increased outflow mass loading are not due to increased stellar clustering,
    but simply due to the higher SN energy budget along with a change in the
    ambient ISM density SNe detonate in.}
    \label{2point_minmass}
\end{figure}
\begin{figure}
    \includegraphics[width=0.5\textwidth]{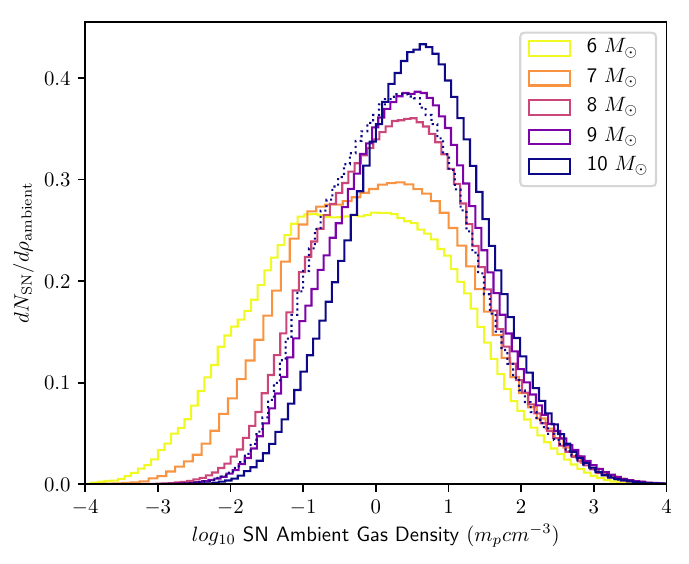}
    \caption{Local ISM density of SN events, as a function of the minimum SN
    progenitor mass.  Unlike what we see for single SN events with various delay
    times in Figure~\ref{sn_rho}, varying the minimum progenitor mass has a
    noticeable change in the typical density in which SNe detonate.  Longer
    injection durations produce a broader distribution of ambient gas densities
    surrounding SN events, with a reduction in the mean density.  As in
    Figure~\ref{sfr_minmass}, we show in the dashed curve a case where the
    energy budget of a $6\Msun$ minimum SN mass is combined with the injection
    duration of a $10\Msun$ minimum progenitor mass.  When this larger energy
    budget is given a shorter injection duration, we see that the ambient
    density about SNe increases.}
    \label{sn_rho_minmass}
\end{figure}

In the previous sections, we have examined how purely numerical choices for
approximating or simplifying the injection timescale of SN energy impact the
effectiveness of the SN feedback at driving galactic outflows and regulating
star formation.  In each of these tests, we have assumed a constant specific SN
energy produced per unit mass of stars formed, $\mathcal{E}_{\rm
SN}=10^{49}\erg\Msun^{-1}$, while varying only the SN delay time $t_0$ and
injection timescale $\tau_{\rm SN}$.  If we consider instead the observational
and theoretical uncertainties in the minimum progenitor mass for SNe, as we have
briefly described in Section~\ref{injection_rates}, it is clear that a different
minimum progenitor mass will change both $\tau_{\rm SN}$ and $\mathcal{E}_{\rm
SN}$.  To that end, we have run a set of 5 galaxies with SN energies and
injection timescales corresponding to the values for a \citet{Chabrier2003} IMF
and a linear approximation of the \citet{Raiteri1996} stellar lifetimes for
minimum progenitor masses of $6-10\Msun$.  The energies and timescales of these
runs are listed in Table~\ref{SN_min_mass_inputs}.  The first SN is determined
by the most massive progenitor, which we fix to $100\Msun$ in all runs, giving
$t_0=3.3\Myr$.  With a constant SN energy of $10^{51}\erg$, the minimum SN
progenitor mass will strongly impact the total SN energy budget
($\mathcal{E}_{\rm SN}$) by changing the number of SN produced per unit stellar
mass formed ($N_{\rm SN}$).  Because we approximate the SN rate as constant
(while in reality, roughly half of SNe occur in the first $\sim25\Myr$ even for
the smallest realistic minimum progenitor mass), assuming a constant luminosity
means that the luminosity for smaller progenitor masses will be slightly lower
than for larger ones: with a $6\Msun$ minimum progenitor, the specific
mechanical luminosity due to SNe is $7.46\times10^{33}\ergs\Msun^{-1}$,
while a $10\Msun$ minimum progenitor mass gives a specific luminosity of
$1.08\times10^{34}\ergs\Msun^{-1}$.  While the instantaneous luminosity may be
lower for a smaller progenitor mass, the longer time over which SNe detonate
produce a larger integrated SN energy budget.  Uncertainties in the minimum SN
progenitor mass produce a much larger difference in the SN energy budget than
uncertainties in the maximum SN progenitor mass or the form of the IMF, as we
showed in figure~\ref{imf_limits}.  While uncertainties in the evolution of SN
remnants and the radiative losses that may occur in an unresolved, magnetized,
turbulent ISM are most likely larger than the $\mathcal{O}(2)$ difference that
the choice of minimum progenitor produces, all models for capturing these
effects must begin with an energy budget set by the range of masses for SN
progenitors.

\begin{table}
    \begin{tabular}{ccc}
        \hline
        Minimum Progenitor Mass & $\mathcal{E}_{\rm SN} (\erg/\Msun)$ &
        $\tau_{\rm SN} (\Myr)$ \\
        \hline
        \hline
        $6\Msun$ & $1.6\times10^{49}$ & 68\\
        $7\Msun$ & $1.3\times10^{49}$ & 48\\
        $8\Msun$ & $1.1\times10^{49}$ & 37\\
        $9\Msun$ & $9.4\times10^{48}$ & 29\\
        $10\Msun$ & $8.2\times10^{48}$ &24\\
        \hline
    \end{tabular}
    \caption{Specific feedback energy and feedback duration for different
    minimum SN progenitor masses derived with a maximum mass of $100\Msun$ and
    the IMF and stellar lifetimes given by equations~\ref{chabrier2003_imf}
    and~\ref{raiteri_lifetimes}.  Increasing the minimum progenitor mass from
    $6\Msun$ to $10\Msun$ not only decreases the overall SN energy budget by
    a factor of $\sim2$, it also decreases the duration of SN injection by
    a factor of $\sim3$.}.
    \label{SN_min_mass_inputs}
\end{table}

The star formation and outflow mass loadings for these different progenitor
masses can be seen in Figure~\ref{sfr_minmass}.  We see here that (as we might
expect), the $\sim2$ times higher SN energy budget and $\sim3$ times longer
injection duration between the smallest progenitor mass $(6\Msun)$ and the
largest progenitor mass $(10\Msun)$ result in a larger reduction in the SFRs
compared to changing the timescale of feedback alone, as we examined in the
previous section.  With a $6\Msun$ minimum progenitor mass, star formation in
the disc is noticeably reduced, with winds driven to an average mass loading of
$\eta=5.5$, more than 2 orders of magnitude larger than the average mass loading
for a $10\Msun$, with $\eta=0.015$.  The dashed curve in
Figure~\ref{sfr_minmass} shows the result of combining the energy budget of a
$6\Msun$ minimum progenitor mass ($1.6\times10^{49}\erg/\Msun$) with the
injection duration of a $10\Msun$ minimum SN progenitor mass ($24\Myr$).  As
this figure shows, the effects of changing the progenitor mass arise from
\textit{both} the increased energy budget and longer SN duration.  As we saw
previously in figure~\ref{sfr_continuous}, the continuous injection of FB energy
can drive stronger outflows compared to instantaneous injection.  Reducing the
injection duration by roughly a third, from $68\Myr$ to $24\Myr$, reduces the
effectiveness of SN feedback at regulating star formation and driving outflows
(although less than if we were to also reduce the feedback energy to
$8.2\times10^{48}\erg/\Msun$).

If we look at the two point correlation function (Figure~\ref{2point_minmass}),
we see again that unlike with instantaneous injection of SN energy, the
increased effectiveness of SN-driven outflows for small minimum progenitor
masses is not associated with an increase in the clustering of star formation.
Instead, these results point to the much simpler effect of higher SN energy
budgets and longer SN injection timescales (resulting in less cooling losses
compared to an instantaneous injection) produce a much stronger driving engine
for powering galactic outflows.  The reduction in cooling losses are clearly
revealed if we examine the ambient ISM densities in which SNe detonate in
Figure~\ref{sn_rho_minmass}.  Here we see the continuous injection of SNe over
$24-68\Myr$ results in a roughly log-normal distribution, without the slight
skewness seen in Figure~\ref{sn_rho}.  In general, we see two major trends: the
width of the distribution of SN events grows with lower minimum progenitor mass,
and the median of the distribution decreases.  For instantaneous injection of
energy, the mean ambient ISM density SN events detonate in is $15-17\hcc$ (for
values of $t_0$ from $0-30\Myr$), while we see mean ambient ISM densities for SN
events of $0.6\hcc$, $1.2\hcc$, $2.0\hcc$, $2.8\hcc$, and $3.9\hcc$ for minimum
progenitor masses of $6\Msun$, $7\Msun$, $8\Msun$, $9\Msun$, and $10\Msun$
respectively. Together with the broader distribution of ambient densities, this
results in $30\%$ of all SN energy being deposited in gas with ambient density
below $0.1\hcc$ with a $6\Msun$ minimum progenitor mass, compared to only $4\%$
for a $10\Msun$ progenitor.  As has been shown in past work
\citep{Sharma2014,Gentry2017}, SNe detonating in the low-density remnants of past
SNe suffer less cooling losses and convert more energy into kinetic motions than
SNe detonating in ``pristine'', gas rich environments.

As the dashed curve in Figure~\ref{sn_rho_minmass} shows, the higher SN energy
budget of a $6\Msun$ minimum progenitor mass ($1.6\times10^{49}\erg/\Msun$),
combined with the shorter SN duration of a $10\Msun$ minimum progenitor
($24\Myr$ vs.  $68\Myr$) does not give the broad, lower-density distribution we
see from setting both the energy budget and duration to the $6\Msun$ values, but
instead is closer to the results of the short-duration injection from a
$9-10\Msun$ minimum progenitor mass.  This explains why the outflow and star
formation rates shown in Figure~\ref{sfr_minmass} does not show as significant
change when the energy budget alone is increased (shown in the dashed curve in
that figure as well).  The importance of the progenitor comes not only from the
change in SN energy budget, but in the duration of injection as well.  A longer
duration allows the most massive SN progenitors, which detonate early, to pre-process the
ISM prior to the detonation of the least massive SN progenitors.  This effect mimics 
early feedback, which has been shown to significantly increase the effectiveness
of SN feedback \citep{Stinson2013,Hopkins2014}.

\section{Discussion}
\label{discussion}
We have seen that the choice of SN delay time $t_0$ and duration $\tau_{\rm SN}$
in simple models for the SN injection rate can have noticeable effects on the
ability of SN feedback at regulating star formation and significant effects on
their effectiveness at driving outflows.  When SNe are injected instantaneously,
with an entire stellar population's SN budget deposited at once, choosing a
delay time $t_0$ of $0\Myr$ vs. $30\Myr$ can change the average SFR by nearly
$50\%$, and the outflow mass loadings by nearly two orders of magnitude, by
greatly increasing the clustering of star formation.  It has been shown that
clustered SNe lose less energy to cooling \citep{Sharma2014,Walch2015a}, produce
higher terminal momenta \citep{Gentry2017}, and ultimately drive more powerful
outflows \citep{Fielding2018}.  \citet{Fielding2017} examined the effect of
clustering directly using idealized, high resolution simulations of isolated
dwarf galaxies.  By keeping the SN rate constant, but varying the fraction of
SNe which occur co-spatially (the clustering fraction $f_{\rm cl}$), they showed
that the mass loading of winds is related to the clustering fraction
$\eta\propto f_{\rm cl}^{1.05}$ \citep[also see Sect. 7.3.4
of][]{Kruijssen2012c}.  The results we have shown here reveal that the SN
injection rate can change the spatial clustering of star formation, resulting in
a similar effect to this.  Clustering has been explicitly employed in the
stochastic feedback model of \citet{DallaVecchia2012}, which artificially
increases the clustering of feedback to overcome numerical losses.  It has also
been seen to be a natural consequence of self-gravity \citep{Martizzi2020},
where the increased clustering seen in the two-point correlation of star
particles has been directly tied to more effective SN driven galactic outflows.
Clustering of star formation has been used in similar simulations to these as a
tool to compare simulations to observations and constrain parameters in models
for feedback and star formation \citep{Buck2019}.  As we show in
Table~\ref{other_SN_rates}, instantaneous injection of SNe is a popular
approximation, and delay times from $0-30\Myr$ have been used in a number of
simulations of cosmological and isolated galaxy evolution.  Our results suggest
that these choices can have comparable effect size to the choice of sub-grid
model for SN feedback.

With a more realistic model for the SN delay time distribution, giving a
continuous injection of energy rather than a simple instantaneous one, we find
that the sensitivity to SN delay time $t_0$ and duration $\tau_{\rm SN}$ is less
clear.  We do not see a significant change in the star formation rate the past
the initial starburst, although we do see somewhat higher averaged outflow mass
loadings with longer delay times.  Interestingly, the most notable change is a
decrease in the ``smoothness'' of the outflows (when $\tau_{SN}>0$), with
order-of-magnitude variations occurring over $\sim100\Myr$ periods.  This
suggests that a more realistic model of SN energy injection may shift towards a
stronger ``fountain'' mode of galactic outflows, where gas is ejected but later
re-accreted.  In \citet{Gentry2020}, the authors found that instantaneous
injection of FB energy significantly reduced the overall momentum injection, but
it is important to consider that the mechanism there (and the mechanism in
\citet{Fielding2018}) may be different than what we show here, as we do not
resolve the cooling radius of most SNe.  Thus, any changes in the cooling
processes on small-scales are not captured by our simulations.

Unfortunately, not all uncertainty can be removed simply by avoiding the
instantaneous injection simplification. Our results in Section~\ref{SN_min_mass}
show that, using a realistic SN delay time distribution designed to fit a 
\citet{Chabrier2003} IMF and a \citet{Raiteri1996} stellar lifetime function,
there is still a major uncertainty in the SN distribution model.  The minimum SN
progenitor mass, as the review by \citet{Smartt2009} details, is uncertain
by at least $1\Msun$.  This means that both the overall SN energy budget, as
well as the SN injection duration, are both uncertain to within roughly
$\sim50\%$.  As Figure~\ref{sfr_minmass} shows, this has a noticeable effect on
the overall SFR of the galaxy and the ability of feedback to
drive outflows.  

An interesting result we have seen as well is the difference in sensitivity to
feedback parameters between the SFR and outflow properties.  Changing the delay
time $t_0$ results in a relatively small ($\sim50\%$) change in SFR, yet drives
a change of nearly two orders of magnitude in outflow rates and mass loadings.
This suggests that the strongest observational tests of SN feedback will be in
examinations of the circumgalactic medium (CGM) and the metal budget of
galaxies.  The CGM is where outflows deposit themselves, and we are now
beginning to see comprehensive observational studies of their mass budgets and
kinematics \citep{Werk2014,Prochaska2017}, which is helping to inform new models
for outflows, the CGM, and galaxy evolution \citep{Voit2019,Keller2020a}.  In
particular, the strong changes in mass loadings suggest that surveys of the
metal loss budget \citep{Kirby2011,Peeples2014,Telford2019} will be a powerful
tool for constraining the detailed parameters of stellar feedback.

We have explored here the impact of three related parameters that must be chosen
for any model of SN feedback: the delay between star formation and the
first SN $(t_0)$, the duration over which SNe detonate $(\tau_{\rm SN})$, and the
total SN energy budget $\mathcal{E}_{\rm SN}$.  These parameters may be chosen as
simplified numerical approximations of the SN delay-time distribution and
energetics, or as a more physically motivated function of the IMF, stellar
lifetimes, and SN progenitor mass function.  This is, however, only a fraction
of the potential sources of uncertainty in the impact of SN feedback.

Of course, SNe are not the only form of stellar feedback, and stellar feedback
is not the only feedback that may influence the evolution of a galaxy.  Energy
released from accretion onto supermassive black holes can power active galactic
nuclei (AGN) which heat the galaxy \citep{McNamara2007} and drive fast outflows
\citep{Morganti2003}.  However, AGN activity occurs in galactic nuclei,
spatially decoupled from the local star forming regions, and will only change
the temporal and spatial clustering of star formation on the scale of the galaxy
itself.  Other forms of feedback from massive stars, however, will have a
similar effect to the parameters we have manipulated here.  Stellar winds
\citep[e.g.][]{Gatto2017}, expanding HII regions \citep[e.g.][]{Franco1990},
ionizing radiation \citep[e.g.][]{Dale2005}, and radiation pressure
\citep[e.g.][]{Krumholz2009} will all act to change the total stellar feedback
energy budget and SN coupling efficiency, as all of these processes begin
immediately after star formation.  These ``early'' feedback mechanisms are
likely responsible for the destruction of molecular clouds
\citep{Kruijssen2019b,Chevance2020,Chevance2022,Kim2021}. It has been shown in previous
simulation studies \citep{Agertz2013,Stinson2013,Hopkins2014} that the
interaction of these multiple feedback mechanisms can result in a nonlinear
change in the effectiveness of SN feedback, more than simply changing the
overall energy budget.  This is most easily explained through two channels: the
pre-processing of ISM gas, lowering the density in which SNe detonate; and the
termination of star formation in dense gas earlier than the $3-5\Myr$ delay
required for the first SN.  However, our results in Section~\ref{continuous}
suggest that an earlier onset of feedback, without any additional energy or
momentum, has a much smaller effect than the use of a realistic delay time
distribution.  How a more complete accounting of the feedback budget might
change our results is a question we leave for future study.

In order to directly compare these universal parameters, rather than the details
of the numerical coupling scheme, we have restricted our simulations to use a
single sub-grid model for SN feedback, the \citet{Kimm2014} mechanical feedback
model that has been widely exploited in the literature.  Previous studies have
shown that choices related to the sub-grid model can have significant impact on
the amount of numerical overcooling \citep{Thacker2000}, SFRs
\citep{Scannapieco2012}, and outflow properties \citep{Rosdahl2017}.  Choices
such as which sub-grid model to use, what mass/length scale to deposit feedback
over, and other model-specific parameters can change the behaviour of SN
feedback to an extent comparable to (or even greater than) the changes in the
model-independent parameters we have studied here.  In particular, the changes
in mass loadings seen when different sub-grid models were compared by
\citet{Rosdahl2017} and \citet{Smith2018} are comparable in magnitude to the
effect of changing $t_0$ from $5\Myr$ to $30\Myr$ with instantaneous injection
of SN energy, or changing the minimum SN progenitor mass from $6\Msun$ to
$10\Msun$.  The most extreme changes in sub-grid model, however, can produce
changes in the SFR significantly larger than we see here, with variations of
nearly an order of magnitude. Unlike these purely numerical choices, however,
the changes in the SN energy budget and duration that arise from different
minimum progenitor masses are driven by uncertainties in the fundamental,
underlying physics of stellar evolution and SNe detonation.  Inferring the
progenitor mass of a SN is a non-trivial process that involves assumptions both
in the stellar evolution model \citep{Williams2018} as well as the distribution
of interstellar and circumstellar gas and dust \citep{Kochanek2012}.  On top of
that, SNe are relatively infrequent events, and only a few dozen events with
identified progenitors have been observed \citep{Smartt2015}.  As long as the
uncertainty in the minimum mass of SN progenitors is large, the energy budget of
SN feedback will have significant uncertainties as well.

We have assumed here that the SN progenitor mass is a continuous range, with all
stars between the minimum and maximum progenitor masses detonating as SNe.
However, this simplifying assumption may not hold. SN progenitors with
initial masses above $\sim18\Msun$ have not been observed \citep{Smartt2015},
suggesting that there may be mechanisms in core collapse that can prevent the
escape of the potential energy that drives SNe.  Simulations of SN core
collapse have suggested that there exist ``islands of explodability'' where the
density profile of the progenitor stellar core allows the neutrino-driven shock
to prevent the fallback of the outer layers of the star.  Stars outside this
mass range may collapse directly to form a black hole without a detectable SN
explosion \citep{Horiuchi2014}.  The regions in which failed SNe occur may
depend not only on mass, but also on metallicity \citep{Heger2003}, rotation
\citep{Hirschi2004}, and binarity \citep{Eldridge2008}.  On top of this,
whether a SN will fail to occur may not be a binary process, with regions
producing SN fractions anywhere from 0 to 1 \citep{Pejcha2015}.

Finally, we have also assumed that all SNe detonate with the same energy,
$10^{51}\erg$, independent of their mass or metallicity.  Just as there may be
regions of the stellar mass-metallicity plane where failed SNe occur, their may
also be regions where subluminous or superluminous SNe occur.
While type II-P SNe are the most common core collapse SN type (as these are the
form that the lowest-mass progenitors are expected to take,
\citealt{Smartt2009}), other core collapse SNe may also be a significant
component of the total SN budget.  SNe associated with gamma ray bursts (GRBs)
may release as much as $2\times10^{54}\erg$ \citep{Woosley2006}, and
sub-luminous SNe have been observed with kinetic energies as low as
$\sim10^{47}\erg$ \citep{Lovegrove2013}.  While the most energetic GRBs are
likely rare, sub-luminous SNe may be as significant a component of the massive
star end sequence as are failed SNe.  The ``islands of explodability'' that are
currently the dominant observational and theoretical paradigm
\citep{Heger2003,Smartt2009,Horiuchi2014} for massive star evolution do not all
produce comparable energies, and the more massive stars may end their lives in
hypernovae that produce 10-100 times as much energy as typical type II-P SN
\citep{Nomoto2011}.  An exploration of how a more physically motivated input
sequence for SN feedback in galaxies that takes into account this more complex
picture of SN energetics is an interesting possible future line of research.

\section{Conclusion}
We have shown here that a few parameters, key to all models for SN feedback in
galaxy simulations, can have a significant impact on the evolution of a galaxy.
A summary of our primary findings are as follows.
\begin{itemize}
    \item Even if the total SN energy budget is kept constant, changing the
        delay between star formation and the first SN ($t_0$) or the duration of
        SNe ($\tau_{\rm SN}$) can have significant effects on the ability of SNe to
        regulate star formation and drive outflows.
    \item Long delays $(t_0>20\Myr)$ between star formation and SN feedback can
        significantly increase the clustering of star formation, and thereby
        increase the effectiveness of SN feedback.
    \item The outflow rate and mass loading of galactic winds is a much more
        sensitive probe of SN feedback timing parameters $t_0$ and $\tau_{\rm
        SN}$ than the overall star formation rate.
    \item Instantaneously injecting SN feedback at or near the time of first SN
        detonation does not significantly change the clustering of star
        formation or the overall star formation rate compared  to injecting
        energy over time, but can reduce the outflow mass loadings of SN-driven
        winds relative to SNe injected over a realistic timescale ($30-40\Myr$).
    \item Continuous injection of SN energy can reduce the ambient ISM density
        SN energy is deposited into, increasing the efficiency of SN feedback at
        regulating star formation and driving outflows as the duration of
        injection increases.
    \item The observational and theoretical uncertainty in the minimum SN
        progenitor mass has a dramatic impact on both the timing and energy
        budget of SN feedback.  Lowering the minimum SN progenitor mass reduces
        the star formation rate, and can increase outflow rates nearly 2 orders
        of magnitude.  
    \item Lower minimum SN progenitor masses injects more SN energy
        over a longer duration.  This of course increases the total feedback
        energy budget, but also results in a larger fraction of SN energy
        deposited in low-density ISM that has been pre-processed by the earliest
        SNe.  This in turn dramatically reduces cooling losses, and increases how
        much kinetic energy each SN event deposits into the ISM.
\end{itemize}

What these results show is that choices in how to represent the SN rate
function can have dramatic effects.  These choices must be made regardless of
the sub-grid method used to capture the effect of SN feedback, and can result in
differences in the star formation and outflow evolution that is comparable to
the differences produced when using different sub-grid models.  Some of these
choices are simple numerical simplifications, such as injecting all SNe
instantaneously after a fixed delay time.  The importance of the delay time
choice can be eliminated by using a more realistic distribution for the SN
lifetimes, giving a delay time distribution that spreads SNe over tens of $\Myr$.
However, even if we use a realistic delay time distribution for SNe, the current
theoretical and observational uncertainties in the minimum mass SN progenitor
translate into uncertainties in the SN injection duration and overall energy.

These uncertainties can also drive large changes in the evolution of the galaxy:
changing the minimum SN progenitor mass from $7\Msun$ to $9\Msun$ results in a
roughly $\sim 25\%$ increase in the overall star formation rate, and a nearly
ten-fold reduction in the outflow mass loadings.  These findings
raise serious doubts as to the level of uncertainty we should have when
interpreting the results of galaxy simulations.  A well-developed literature has
shown that the choice of SN sub-grid models can have dramatic impact on the
evolution of the galaxy.  We have shown here that there are choices below that
sub-grid, in the injection timescales and SN budget, that can also dramatically
impact the galaxy.  There is clearly still significant uncertainty in the
effectiveness of SN feedback at regulating star formation and driving outflows
in galaxy simulations.  Taking together our results, we recommend future
simulations use at the least a constant injection of SN energy, rather than an
instantaneous approximation.  Along with this, whenever possible, simulations
should bound their results by running examples with integrated SN energies and
durations for minimum progenitor masses of $7\Msun$ and $9\Msun$: $\tau_{\rm
SN}=48\Myr$ with $\mathcal{E}_{\rm SN}=1.3\times10^{49}\erg\Msun^{-1}$ for a
$7\Msun$ minimum progenitor, and $\tau_{\rm SN}=29\Myr$ with $\mathcal{E}_{\rm
SN}=9.4\times10^{48}\erg\Msun^{-1}$ for a $9\Msun$ minimum progenitor.  While
this will result in a modest increase in computational expense, it will allow
one to quantify the impact that uncertainties in underlying SN physics have on
all predictions made by these future simulations.

\label{conclusion}

\section*{Acknowledgements}
We thank Volker Springel for allowing us access to {\sc Arepo}. The analysis was
performed using astroML (\texttt{http://astroml.org}, \citealt{astroML}) and
pynbody (\texttt{http://pynbody.github.io/}, \citealt{pynbody}).   BWK
acknowledges funding in the form of a Postdoctoral Research Fellowship from the
Alexander von Humboldt Stiftung. BWK and JMDK gratefully acknowledge funding
from the European Research Council (ERC) under the European Union's Horizon 2020
research and innovation programme via the ERC Starting Grant MUSTANG (grant
agreement number 714907).  JMDK gratefully acknowledges funding from the
Deutsche Forschungsgemeinschaft (DFG, German Research Foundation) through an
Emmy Noether Research Group (grant number KR4801/1-1) and the DFG Sachbeihilfe
(grant number KR4801/2-1).

\section*{Data Availability Statement}
The simulation data used in this paper will be shared on reasonable request to
the corresponding author.

\bibliographystyle{mnras}
\bibliography{references}
\appendix

\section{Robustness to Stochasticity} 
\label{stoch}
Recent studies \citep{Keller2019,Genel2019} have shown that galaxy evolution
shows non-trivial stochasticity, owing to the chaotic nature of the N-body
problem that describes said evolution.  Here we attempt to quantify this effect in our
simulations, in order to ensure that our results are indeed due to the changes
in clustered star formation and wind driving, rather than random run-to-run
variations.  In order to do this, we re-simulate 8 instances of our
instantaneous SN injection cases with $0\Myr$, $5\Myr$, and $30\Myr$ delays.  In
order to seed variations between the 8 runs, we use a different
number of cores for each, thus introducing differences in the domain
decomposition and communication patterns at the level of floating-point roundoff.

\begin{figure}
    \includegraphics[width=0.5\textwidth]{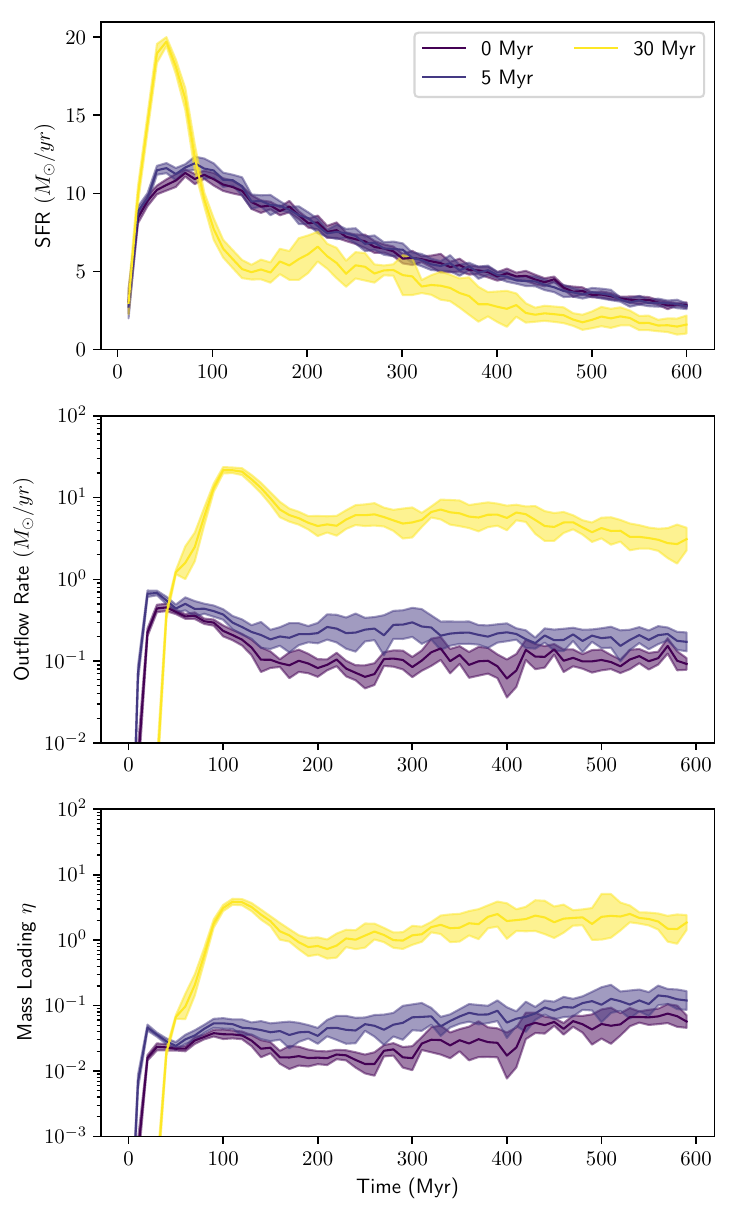}
    \caption{Median values (solid lines) and $\pm1\sigma$ variance (shaded
    regions) for the star formation rate (top panel), mass outflow rate (middle
    panel), and outflow mass loading (bottom panel) for sets of 8 variance
    re-simulations with SN delay times of  $0\Myr$, $5\Myr$, and $30\Myr$.  As
    can be seen, the difference in SFR between the two shorter
    delay times are mostly within the run-to-run uncertainty after $300\Myr$,
    but the difference between these and the longer delay are distinct save for
    a short bursty period from $200-400\Myr$.  The outflow rates and mass
    loadings, however, are significantly different, even despite the larger
    variances for each case.}
    \label{sfr_variance}
\end{figure}

As can be seen in Figure~\ref{sfr_variance}, the variation in star formation
rates is large enough that the delays of $0\Myr$ and $5\Myr$ are nearly
indistinguishable $300\Myr$ after the beginning of the run and the initial
starburst.  The drop in SFR for the longer delay,
however, is significant enough that the difference is nearly always outside the
run-to-run variation.  For the outflow rates and mass loading, there is both a
larger difference between the median values for each delay time, and a larger
run-to-run variation.  The differences between the delay times, though, are
clearly evident, and we can therefore say with confidence that increased 
wind launching efficiency that occurs for longer SN delay times is a real
effect, and not simply a fluke produced by random variance.

\section{Sensitivity to Initial Starburst} 
\label{starburst}
Isolated galaxy simulations frequently suffer from transient bursts in star
formation shortly after the simulation begins.  We see this in the simulations
presented here, and similar effects can be seen in the isolated galaxies from
other studies (e.g. \citealt{Keller2014}, \citealt{Rosdahl2017},
\citealt{Valentini2019}).  In order to verify that the effects we see here are
not simply due to the different strength of the initial starburst, we ran two
additional experiments.  In figure~\ref{sfr}, it is clear that longer SN delays
produce a stronger initial burst in star formation.  Thus, it is possible that
the changes in the star formation and outflow rates are a result of this
initial burst re-shaping the disc ISM.  To verify this is not the case, we took
the two extremal simulations ($t_0=0\Myr$ and $t_0=30\Myr$) and re-started them
at the $300\Myr$ point with swapped values for $t_0$.  This gives us two
additional runs to compare: a run which spends the first $300\Myr$ with
$t_0=0\Myr$ and the final $300\Myr$ with $t_0=30\Myr$, and a run which spends
the first $300\Myr$ with $t_0=30\Myr$ and the final $300\Myr$ with $t_0=0\Myr$.

\begin{figure}
    \includegraphics[width=0.5\textwidth]{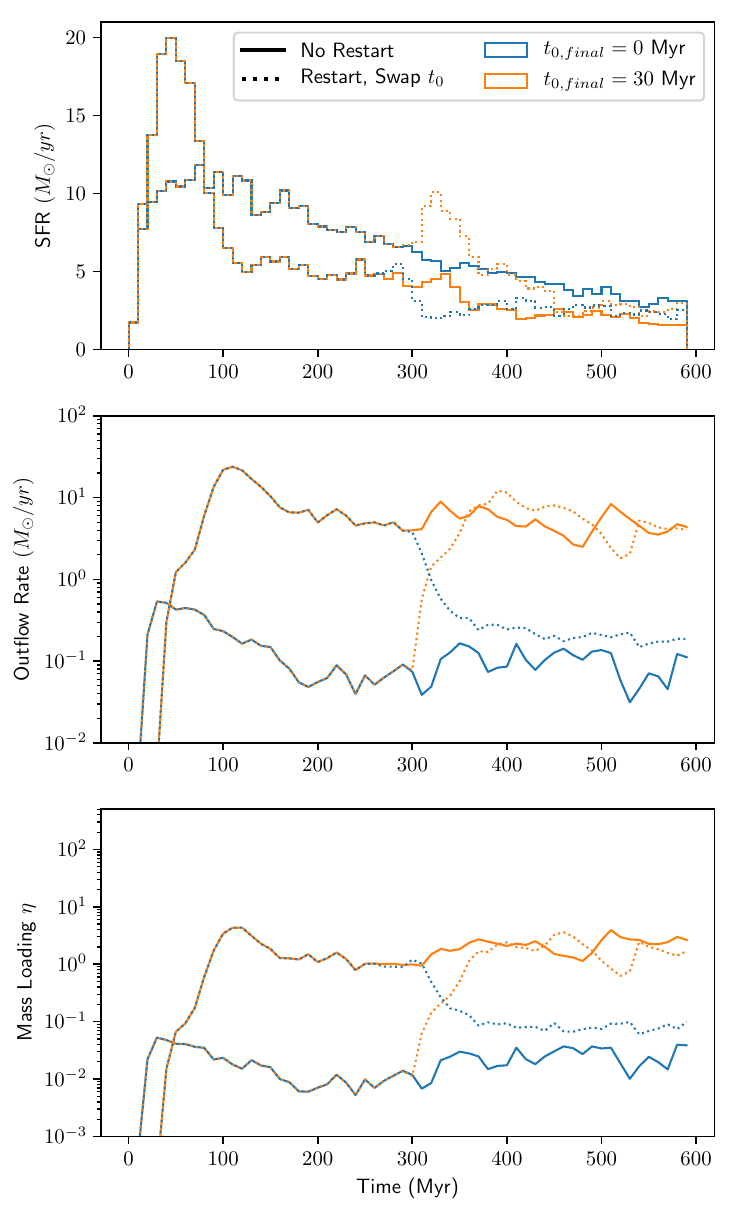}
    \caption{Comparison of the star formation rate (top panel), outflow rate
    (middle panel), and mass loading (bottom panel) of four permutations of the
    SN delay time $t_0$.  The two solid lines are run for all $600\Myr$ with
    constant $t_0$, while the dashed lines swap the value of $t_0$ from $0\Myr$
    to $30\Myr$ or vice versa at $300\Myr$.  As is clear, the final star
    formation rates, outflow rates, and mass loadings are set by the value of
    $t_0$ after the initial star burst, not the value during the star burst.}
    \label{sfr_restart}
\end{figure}

If the strength of the initial burst in star formation is the cause of the
differences between the $t_0=0\Myr$ and $t_0=30\Myr$ runs we see in
section~\ref{delay_time}, then changing the value of $t_0$ after this burst has
completed shouldn't change the evolution of the galaxy.  However, if the initial
burst is not important, and the on-going star formation (and its clustering) is
what establishes the changes we have seen, then changing the value of $t_0$
should push the SFR, outflow rates, and mass loadings towards the values they
would have had if $t_0$ was set to the final value since the beginning of the
simulation.  In other words, changing the value of $t_0$ from $0\Myr$ to
$30\Myr$ should slightly decrease the star formation rate and significantly increase the
outflow rate, while doing the opposite change should produce the opposite
result.  As we show in figure~\ref{sfr_restart}, this is exactly what occurs.
This demonstrates clearly that the effects we have examined are not simply the
aftereffects of the initial burst in star formation.  

\begin{figure}
    \includegraphics[width=0.5\textwidth]{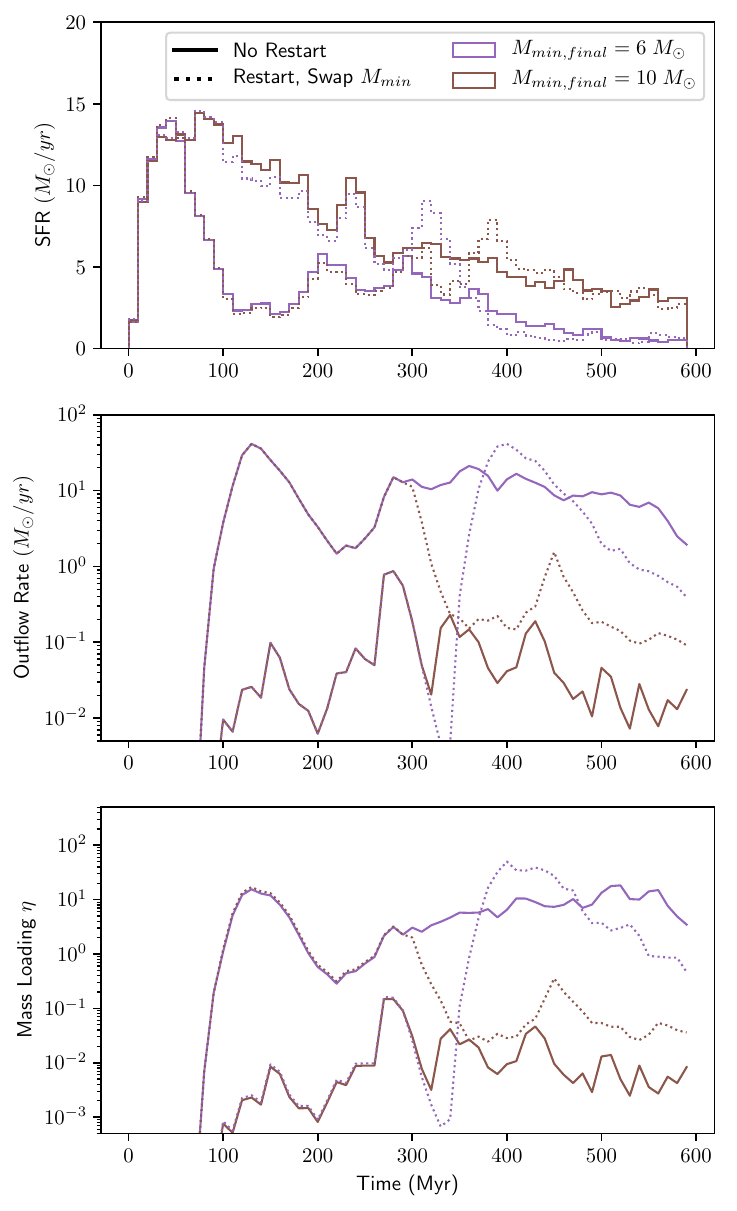}
    \caption{Comparison of the star formation rate (top panel), outflow rate
    (middle panel), and mass loading (bottom panel) of four permutations of the
    minimum SN progenitor mass $M_{min}$.  The two solid lines are run for the
    entire simulation with constant minimum SN progenitor mass ($6\Msun$ in
    purple, $10\Msun$ in brown).  The dashed lines swap the value of the minimum
    progenitor mass at 300 Myr.  As this shows, the initial burst in star
    formation from the initial conditions does not set the differences we see in
    the star formation rates, outflow rates, and mass loadings.  These
    differences are instead set by the change the strength of SN feedback for
    different progenitor masses.}
    \label{sfr_minmass_restart}
\end{figure}

We also verify that the initial burst of star formation as the disc settles is
not dominating the changes we see in varying the minimum SN progenitor mass.  As
with the previous experiment, we compare simulations that use $6\Msun$ and
$10\Msun$ minimum SN progenitor masses for their entire $600\Myr$ duration to
runs that swap the minimum progenitor mass at $300\Myr$.  In
figure~\ref{sfr_minmass_restart}, we show that the simulations which swap the minimum
progenitor mass rapidly diverge from the simulations which do not.  While they
do not settle to the exactly same star formation and outflow rates, the
simulations that swap the minimum mass behave similarly to the simulations which
do not, but share the same final minimum SN progenitor mass.  This again
verifies that the difference we see in the star formation and outflow rates are
not simply an artifact of the initial star burst.
\end{document}